\documentclass{aa}
\usepackage{txfonts}
 \usepackage{ifpdf}
 \ifpdf
 \usepackage[pdftex]{epsfig}
 \else
 \usepackage[dvips]{epsfig}
 \fi
\usepackage{graphicx}
\usepackage{natbib}
\usepackage{lscape}
\usepackage{float}

\begin{document}
\bibpunct{(}{)}{;}{a}{}{,} 
\titlerunning{Evolution in the diagnostic diagrams}
\title{Galaxy evolution across the optical emission-line diagnostic diagrams?}

\author{M. Vitale\inst{1,2}
  \and L. Fuhrmann\inst{2}
  \and M. Garc\'{i}a-Mar\'{i}n\inst{1}
  \and A. Eckart\inst{1,2}
  \and J. Zuther\inst{2,3}
  \and A. M. Hopkins\inst{4}
}
 

\offprints{Mariangela Vitale, \email{vitale@ph1.uni-koeln.de}}

\institute{I. Physikalisches Institut, Universit\"at zu K\"oln, Z\"ulpicher Strasse 77, 50937 K\"oln, Germany
  \and Max-Planck Instutut f\"ur Radioastronomie, Auf dem H\"ugel 69, 53121 Bonn, Germany
  \and Bonn-Cologne Graduate School of Physics and Astronomy, Universit\"at zu K\"oln, Z\"ulpicher Strasse 77, 50937 K\"oln, Germany
  \and Australian Astronomical Observatory, P.O. Box 915, North Ryde, NSW 1670, Australia}                                                                                                 


\abstract {The discovery of the $M-\sigma$ relation, the local galaxy bimodality, and the link between black-hole and host-galaxy properties, have raised the question whether Active Galactic Nuclei (AGN) play a role in galaxy evolution. AGN feedback is one of the biggest observational challenges of modern extragalactic astrophysics. Several theoretical models implement AGN feedback to explain the observed galaxy luminosity function, and possibly the color and morphological transformation of spiral galaxies into passive ellipticals.}{To understand the importance of AGN feedback, a study of the AGN populations in the radio-optical domain is crucial. A mass sequence linking star-forming galaxies and AGN has been already noted in previous works, and it is now investigated as possible evolutionary sequence.}{We observed a sample of $119$ intermediate-redshift ($0.04 \leq z<0.4$) SDSS-FIRST radio emitters with the Effelsberg 100-m telescope at $4.85$ and $10.45$ GHz and obtained spectral indices. The sample 
includes star-forming galaxies, composite galaxies (with mixed contribution to line emission from star formation and AGN activity), Seyferts and Low Ionization Narrow Emission Region (LINER) galaxies. With these sources we search for possible evidence of spectral evolution, and a link between optical and radio emission in intermediate-redshift galaxies.}{We find indications of spectral index flattening in high-metallicity star-forming galaxies, composite galaxies, and Seyferts. This ``flattening sequence'' along the [NII]-based emission-line diagnostic diagram is consistent with the hardening of galaxy ionizing field, due to nuclear activity. After combining our data with FIRST measurements at $1.4$ GHz, we find that the three-point radio spectra of Seyferts and LINERs show substantial differences, attributable to small radio core components and larger (arcsecond sized) jet/lobe components, respectively. A visual inspection of FIRST images seem to confirm this hypothesis.}{Galaxies 
along this 
sequence are hypothesized to be transitioning from the active star-forming galaxies (blue cloud) to the passive elliptical galaxies (red sequence). This supports the suggestion that AGN play a role in shutting down star-formation, and allow the transition from one galaxy class to the other.}

\keywords{Galaxies: active -- Galaxies: evolution -- Galaxies: starburst -- Radio:continuum}
\maketitle

\section{Introduction}
The observationally established correlations between the mass of the black hole (BH) and the properties of the bulge of its host galaxy \citep{KormendyRichstone1995,Magorrian1998,Gebhardt2000,Ferrarese2000,Tremaine2002,Marconi2003,Heckman2004,Aller2007,Hopkins2007,FeoliMancini2009, Gueltekin2009} support the hypothesis of coevolution between BH and surrounding galaxy. A tight link between black hole growth and evolution of the host is expected, considering that BH accretion and star formation are fueled by the same material \citep{Trump2012}. Observational evidence for coeval star formation and black hole growth are found in the star-formation histories of Active Galactic Nuclei (AGN) hosts \citep{HeckmanKauffmann2006}.

Galaxy color bimodality \citep[e.g.][]{Baldry2004} leads to the expectation that galaxy color may evolve with time, followed by a morphological evolution \citep{Kauffmann2003a}. When molecular gas is available, the galaxy appears blue \citep{Baldry2004}. It is generally accepted that massive ellipticals form via mergers of gas-rich disk galaxies \citep[e.g.][]{Toomre1972}. The latter provide a significant amount of gas to fuel star formation. However, models for galaxy evolution that explain the subsequent passive phase via gas consumption, produce galaxies that are too luminous and blue \citep{Croton2006,Bower2006}. In between the blue distribution of star-forming objects (blue cloud) and the red distribution of passively evolving galaxies (red sequence) lies a group of galaxies showing intermediate properties (green valley). The latter also show intermediate morphology, i.e. they present large bulges with a disk component. The relatively high number of AGNs that have been found to reside in green valley 
galaxies \citep{Nandra2007,Salim2007,Georgakakis2008,Silverman2008,Gabor2009,Schawinski2009,Hickox2009} may connect the black hole activity with the transitional phase from the blue cloud to the red sequence. This has been construed as AGN-feedback \citep{SilkRees1998,Benson2003,DiMatteo2005,Cattaneo2005,KawataGibson2005,Kaviraj2007,Khalatyan2008}, where plasma jets may sweep away or heat up the gas reservoir of galaxies. This would remove the fuel for star formation, or prevent it from cooling down to form new stars \citep{DekelSilk1986,Benson2003}. In particular, the existence of two different AGN-feedback modes has been suggested. In the \textit{quasar} or \textit{high-excitation mode}, a fraction of the gas accreted onto a BH during a major merger is isotropically injected back to the gas reservoir of the host galaxy as thermal energy. According to simulations \citep{Springel2005,Narayanan2006,Li2007,Narayanan2008} this may 
be the case for quasar-driven powerful outflows that quench the star formation on short time scales \citep{Hamann2002,Maiolino2012}. This phase is also associated with vigorous star formation at kpc scales. It is now observationally established that in some cases the most luminous radio-loud AGNs are associated with enhanced star formation activity \citep{Ivison2012,Norris2012}.
\citet{Tadhunter2007} state that the percentage of powerful radio loud AGNs with significant recent starburst activity is about $20-30\%$. Cases of radio-loud AGNs with powerful recent star formation can be found in \citet{Hes1995,Tadhunter2002,Wills2002,Wills2004,Tadhunter2007}. Towards the end of the quasar phase, the AGN heats or mechanically disrupts the surrounding gas, and star formation is terminated. The host is then visible as an early-type galaxy whose spectrum is dominated by older generations of stars with redder colors. Theoretical models also include quasar feedback as a way to quench star formation in the host and create red passively-evolving products \citep{Granato2004,Springel2005,Hopkins2006a}. The second AGN-feedback mode is the so-called \textit{radio mode} or \textit{low-excitation mode} \citep{Croton2006,Bower2006,Khalatyan2008}. It is thought to occur at later times and over a more extended period of time, liberating a smaller fraction of energy into the ISM and preventing 
cooling 
flows. The radio mode is linked to 
radio AGN outflows in already massive, red galaxies. The destruction of the molecular gas reservoir may in turn terminate the growth of the black hole as material for further accretion is no longer available.

The observational finding that AGN can also trigger star formation activity points to a more complex interplay between the two \citep{Khalatyan2008,Cattaneo2009,Harrison2012, IshibashiFabian2012,Zinn2013}. Enhanced star formation can be triggered by strong jets, which cause shocks and induce turbulence \citep{Klamer2004,vanBreugel2004,Gaibler2012,IshibashiFabian2012}.
The question to what extent AGN and host are related to each other still needs to be answered. Strong and direct evidence of AGN-feedback in action is still missing.

Radio galaxies are active galaxies that are luminous at radio frequencies, and are often chosen for AGN-feedaback studies because jets and lobes are more easily observed. They can be divided into two main groups. The Fanaroff-Riley type I \citep[FR I,][]{Fanaroff1974} sources become fainter as one approaches the outer points of the lobes. Radio spectra are steep, indicating that the radiating particles have aged the most. Fanaroff-Riley type II (FR II) sources show the oppsite trend, i.e. hotspots in their lobes at great distance from the center. The majority of FR I radio galaxies show very weak or completely absent optical emission lines. Those are referred to as low-excitation systems \citep{Hardcastle2006}, and they are mostly found in elliptical galaxies with little ongoing star formation \citep{LedlowOwen1995,Govoni2000}. If optical spectroscopic information is available, low-excitation systems are generally classified as low-ionization nuclear emission-line regions \citep[LINERs,][]{Heckman1980}. 
Conversely, the most powerful, high-redshift FR II radio galaxies have, in most cases, strong low-ionization emission lines typical of AGN spectra \citep{Wierzbowska2011} and the hosts show peculiar optical morphologies like tails, bridges, and shells \citep{SmithHeckman1989,Zirbel1996}, indicating recent galaxy-galaxy interactions, and bluer colors compared to giant ellipticals. \citet{Schawinski2007} explain the bluer colors of early-type galaxies hosting AGN as ``rejuvenation'', i.e. star formation triggered by the AGN. FR Is and IIs are usually associated with elliptical hosts \citep{Martel1999}, but exceptions are found \citep[e.g.,][]{Ledlow1999}. While FR IIs show post-merger features like tails and bridges, and are hypothesized to result from the encounter of gas-rich parent galaxies, some FR Is might be the product of non-gas rich galaxies encounters \citep{Gonzalez-Serrano1993,Colina1995,Martel1999}. 

It has been established that powerful FR IIs show a strong correlation between their radio luminosity and their optical emission-line luminosity \citep{Baum1995}, suggesting that both the optical and the radio emission are connected to each other and might arise from the same physical process.
Some attempts to study the correlation between AGN activity and evolution of host properties in the radio-optical regime have been already done in the past by using the large SDSS sample combined with the Faint Images of the Radio Sky at Twenty-Centimiters \citep[FIRST;][]{Becker1995}, and the NRAO VLA Sky Survey \citep[NVSS;][]{Condon1998}, or previous radio samples \citep{Ivezic2002,Obric2006,Buttiglione2010,Best2005,Vitale2012}. \citet{Vitale2012}, using a cross-matched SDSS-FIRST sample, find that the relative number of AGNs decreases with the ratio between radio luminosity at $1.4$ GHz ($L_{\rm r}$) and luminosity of the H$\alpha$ line ($L_{\rm {H\alpha}}$). In particular, the radio luminosity increases from star-forming galaxies to composites (galaxies where contributions to line emission come from both AGN and star formation), Seyfert and LINER galaxies, while the H$\alpha$ luminosity decreases and reaches its minimum in LINERs. This can be interpreted as optical line emission tracing recent star-
formation in 
relatively young galaxies, while radio emission appears at a later stage of galaxy evolution and is mostly detected in massive and red ellipticals. This fact is consistent with the differences in the radio luminosity functions of starburst and AGNs \citep{MachalskiGodlowski2000,Sadler2002}.

Low-ionization emission-line diagnostic diagrams  \citep{Baldwin1981,Veilleux1987,Dopita2002,Kewley2003,Lamareille2004,Groves2004,Groves2004b,Lamareille2010} are largely used to classify galaxies according to the main contributions to line emission - photoionization by hot stars and/or Supernovae (SNe) driven shocks from HII regions, photoionization by disk accretion or fast shocks from AGN, or both. Diagnostic diagrams make use of emission line ratios whose strength is a function of the ionization parameter $U$, and the metallicity. In general, AGNs differ from star-forming galaxies because their ionizing fields have a higher energy component, and produce strong [N\,{\sc{ii}}], [S\,{\sc{ii}}], and [O\,{\sc{i}}] lines.

To understand the importance of AGN feedback in the current scenarios of galaxy evolution, and the connection between nuclear activity and change in the physical properties of the host, a study of the AGN populations in the radio-optical domain is crucial. In the scenario where powerful (predominantly high-excitation) and weak (predominantly low-excitation) radio AGNs represent different (i.e., earlier and later, respectively) stages of the blue-to-red galaxy evolution \citep{Smolcic2009}, radio AGN activity is a strong function of host galaxy properties. Radio activity can therefore be linked to different stages of massive galaxy formation \citep[see also][for a similar study in the optical domain]{Schawinski2009}. The indication of a sequence of increasing stellar mass from star-forming galaxies to AGNs as found in \citet{Vitale2013}, the increase of the $L_{\rm r}/L_{\rm {H\alpha}}$ ratio along the same \citep{Vitale2012}, and the color and morphological transformation, all point to a possible AGN-
triggered evolution 
that needs to be tested.

This paper aims at studying galaxy spectral evolution in a sample of bright intermediate-redshift radio emitters with detected optical counterparts. We performed observations at $4.85$ and $10.45$ GHz with the Effelsberg 100-m telescope of a sample of $119$ radio emitters to study the spectral properties of star-forming, composite, Seyfert, and LINER galaxies. We search for AGN-feedback signatures in the ``transiting'' or composite galaxies, which may represent the intermediate stages of galaxy evolution.

This paper is organized as follows: In Sec. 2 we introduce the sample selection and available archival data, and in Sec. 3 the radio observations and data reduction are presented. Data analysis and results are described in Sec. 4. Sec. 5 contains the discussion, and Sec. 6 the main findings and conclusions.

\section{Optical and radio samples}
\subsection{Sloan Digital Sky Survey}
The Sloan Digital Sky Survey (SDSS) is a photometric and spectroscopic survey that covers one-quarter of the celestial sphere in the north Galactic cap \citep{York2000,Stoughton2002}. Observations were performed using a $2.5$ m wide-field optical telescope. The spectra have an instrumental velocity resolution $\sigma$ of $\sim 65$ km/s and the spectral coverage is $3\,800-9\,200~$\AA. The measured galaxies have a median redshift of $z\sim0.1$. Spectra were taken with $3\arcsec$ diameter fibers ($5.7$ kpc at $z\sim0.1$), which makes the sample sensitive to aperture effects (see Sect. 2.4). Low redshift objects are likely to be dominated by nuclear emission \citep[e.g.,][]{Kewley2003}.\newline
The Max-Planck-Institute for Astrophysics (MPA)-Johns Hopkins University (JHU) Data Release 7 (DR7) of spectral measurements\footnote{http://www.mpa-garching.mpg.de/SDSS/DR7/} contains the derived galaxy properties from the MPA-JHU emission line analysis for the $\sim10^6$ objects in the SDSS DR7 \citep{Abazajian2009}. Stellar population synthesis spectra \citep[updated][]{BruzualCharlot2003} have been used for the stellar continuum subtraction. The MPA-JHU contains, in particular, the emission-line measurements of the low-ionization lines that are used in diagnostic diagrams to separate star-forming galaxies from the AGN families (Seyferts and LINERs, or high- and low-excitation systems). Optical images are available as well. Total stellar masses are also provided and estimated by fitting broad-band spectral energy distributions (SEDs) with stellar population models. 

\subsection{Faint Images of the Radio Sky at Twenty-centimeters survey}
The Faint Images of the Radio Sky at Twenty-centimeters survey \citep[FIRST;][]{Becker1995} has used the Very Large Array (VLA) in the B-array configuration to produce a map of the $1.4$ GHz sky emission with a beam size of $5\farcs4$ and a rms sensitivity of about $0.15$ mJy/beam. The survey covers an area of about $10\,000~\rm deg^2$ in the north Galactic cap, corresponding to the sky regions investigated by SDSS, and observed $\sim10^6$ sources. At the $1$ mJy source detection threshold, about one third of the FIRST sources show resolved structures on scales of $2\arcsec-30\arcsec$ \citep{Ivezic2002}. The FIRST catalog\footnote{http://sundog.stsci.edu/first/catalogs/readme.html} provides information on the continuum flux density peak and the integrated flux density at $1.4$ GHz, which allows the separation of resolved from unresolved sources. FIRST sources have flux density measurement errors smaller than $8$\%.
Radio images are available for each source.

\subsection{SDSS-FIRST cross-identification}
We performed a cross-identification of the SDSS DR7 and the FIRST catalog to obtain a large optical-radio sample of mostly active, metal rich galaxies \citep[see also][for details]{Vitale2012}. For generating the cross-matched SDSS-FIRST sample, we used the matching results provided by the SDSS DR7 via Casjobs \citep{OMullane2005}, based on a matching radius of $1\arcsec$. The resulting sample contains $37\,488$ radio emitters which corresponds to $4$\% of the SDSS sample. Within the detection limits of the FIRST and SDSS surveys, there are radio emitters without optical counterparts. \citet{Best2005} found that radio galaxies extracted from the main spectroscopic sample of the SDSS reside in very massive early-type galaxies, with weak or undetectable optical emission lines. Some potentially real cross-matched objects are not identified by the matching procedure, however, due to positional offsets between the location of the radio emission with respect to the optical emission, especially in case of extended 
sources. From here and throughout the paper, we call the cross-matched SDSS-FIRST sample the ``parent sample''.
   
\begin{figure} 
  \centering
  \includegraphics[width=9cm]{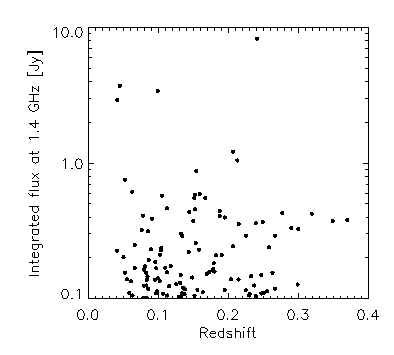}
  \caption{FIRST radio fluxes ($1.4$ GHz) versus redshift for the $119$ galaxies observed with the Effelsberg telescope.}
 \label{radio_fluxes_vs_z}
 \end{figure}

\subsection{Effelsberg sample selection}
From the parent sample we have selected a subsample of radio-bright galaxies ($F_{1.4}\geq 100$ mJy, where $F_{1.4}$ is the integrated flux density at $1.4$ GHz) to obtain spectral information at two different radio frequencies than FIRST. Observations were performed with the Effelsberg 100-m single dish telescope (see Sect. 3). 

The $F_{1.4}\geq 100$ mJy selection resulted in a sample of $263$ sources, among which $175$ exhibit SDSS emission line measurements used in optical diagnostic diagrams. From this initial sample, we excluded: 
\begin{itemize}
 \item all line emitters with $z < 0.04$ to avoid aperture effects \citep[see][]{Kewley2003};
 \item one high redshift ($z>0.4$) galaxy; 
 \item three sources with broad lines;
 \item all sources with signal to noise (S/N) $<3$ on the equivalenth width of the emission lines involved in the diagnostic diagrams.
\end{itemize}

\begin{figure} 
  \centering
  \includegraphics[width=9cm]{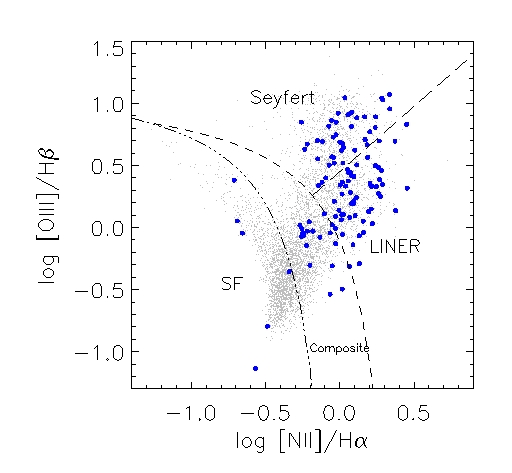}
  \caption{[NII]-based diagnostic diagrams of the parent (grey) and Effelsberg (blue) samples. Demarcation lines were derived by \citet{Kewley2001} (dashed) to set an upper limit for the position of star-forming galaxies, and by \citet{Kauffmann2003a} (three-point dashed) to trace the observed lower left branch (purely star-forming galaxies) more closely. The dividing line between Seyferts and LINERs (long dashed) was set by \citet{Schawinski2007}.}
 \label{classification_dd_NEW}
 \end{figure}
 
Galaxies of the Effelsberg sample are in the redshift range $0.04 \leq z < 0.4$. The lower cut is because local galaxies have angular sizes larger than the size of the fiber used for the SDSS survey. Assuming good centering, emission from the outer regions of the galaxies is not measured. Consequently, a redshift lower limit of $z \geq 0.04$ is required \citep{Kewley2003} where the SDSS fiber includes on average $>20\%$ of the galaxy area. The redshift upper limit is due to the reliance on emission-line diagnostic diagrams, requiring [N\,{\sc{ii}}] and H$\alpha$ in the observable spectral window, hence limiting the sample to $z \lesssim 0.4$. The [OII]/H$\beta$ versus [O\,{\sc{iii}}]/H$\beta$ diagram \citep{Tresse1996,Rola1997,Lamareille2004,Lamareille2010} represents a higher-redshift option to the more classic diagnostic diagram, although there is only one sources in our sample at $z = 0.54$. The use of these additional diagnostics does not add substantially to our analysis, and the source has been 
excluded from the sample. The redshift distribution is illustrated in Fig. \ref{radio_fluxes_vs_z}. There is no obvious trend between integrated flux at $1.4$ GHz and redshift.

The cut in S/N on the equivalenth width of the emission lines involved in the diagnostic diagrams leads to a sample of $122$ galaxies in the [N\,{\sc{ii}}]-based diagram (S/N $>3$ for [N\,{\sc{ii}}], [O\,{\sc{iii}}], H$\alpha$ and H$\beta$ lines), $103$ galaxies in the [S\,{\sc{ii}}]-based diagram (S/N $>3$ for [S\,{\sc{ii}}] doublet, O\,{\sc{iii}}, H$\alpha$ and H$\beta$ lines), and $103$ galaxies in the [O\,{\sc{i}}]-based diagram (S/N $>3$ for [O\,{\sc{i}}], O\,{\sc{iii}}, H$\alpha$ and H$\beta$ lines). The [N\,{\sc{ii}}]-based diagram counts the highest number of sources and is used as main tool to show our results. 
With the Effelsberg telescope we observed $119$ (out of $122$) [N\,{\sc{ii}}]-diagram classified high S/N sources (see Sect. 3.1). Sources are listed in Table \ref{all_info} of the Appendix, where also additional diagnostic diagrams are shown. The sample is composed of $6$ star-forming galaxies (SFGs, $5$\% of the sample), $19$ composites ($16$\%), $42$ Seyferts ($35$\%), and $52$ LINERs ($44$\%). Fig. \ref{classification_dd_NEW} shows how sources are distributed in the diagnostic diagram. The underlying grey distribution represents the SDSS-FIRST parent sample.

The completeness of the Effelsberg sample is related to the completeness of SDSS and FIRST. SDSS is regarded as a complete sample (magnitude limited, but including all galaxy types), while FIRST is surface-brightness (Jy beam$^{-1}$) limited \citep{Condon1998b}. For more information on these samples, see \citet{Vitale2012}. By selection, all galaxies in the sample are line emitters (with available SDSS measurements). Seyfert galaxies usually have bright and well detectable emission lines, while LINERs exhibit much weaker emission lines. By applying a S/N cut on the line fluxes, we likely exclude some of the optically-weak but radio-strong emitters from the sample. In particular, we could be excluding galaxies such as the so-called ``fake-LINERs'' and ``retired galaxies'' \citep{Stasinska2008,CidFernandes2010,CidFernandes2011,Vitale2012}, which exhibit low H$\alpha$ equivalent width and are possibly powered by old post-asymptotic giant branch (AGB) stars. 
The Effelsberg sample, selecting all cross-matched SDSS-FIRST radio emitters that have an integrated flux at $1.4$ GHz $\geq 100$ mJy, and a redshift $0.04\leq z<0.4$, contains many AGNs and metal-rich star-forming galaxies \citep[see][Fig.1]{Vitale2012}. This is because, in general, high radio-flux sources are found to be powerful AGN, while the fainter sources include many more normal (non active) or star-forming galaxies. Consequently, we note that our sample is not complete in a statistical sense, and findings are not representative of the whole optical-radio galaxy population.

\section{Effelsberg}
\subsection{Observations}
Observations at the Effelsberg 100-m telescope were performed between 
February and October 2013 in a total of $7$ single observing sessions. 
Each source was observed at 4.85\ GHz (6\,cm) and 10.45\,GHz 
(2.8\,cm) with multi-feed heterodyne receivers mounted 
in the secondary focus to derive radio spectral indices
from (quasi-) simultaneous observations (guaranteeing measurements free 
of source variability). The flux density measurements at each frequency 
were performed using cross-scans in Azimuth and Elevation direction 
as well as ``beam switch''. The cross-scans allow the correction of 
small pointing offsets and the detection of possible spatial extension 
or cases of confusion from field sources. The beam switch allows 
real-time sky subtraction and removes most of the atmospheric 
variations. All sources in the sample were sufficiently strong 
to facilitate cross-scans with typically 4 to 12 sub-scans (each 
with a length of $3.5$ times the telescope beam) depending on the 
source brightness at the given frequency. During each session 
frequent focus and calibration measurements were performed using 
the primary calibrators 3C\,286, 3C\,295 and NGC\,7027. 
Three weak galaxies were not observable during our session at Effelsberg due to bad weather conditions.   
\begin{figure*} 
  \centering
  \includegraphics[width=19cm]{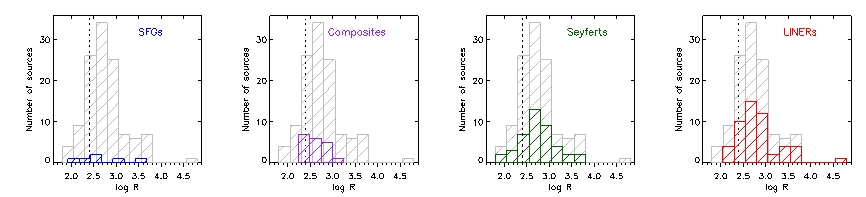}
  \caption{Radio-loudness distribution for each spectral kind as classified by the [NII]-based diagram. SFGs are in blue, composites in purple, Seyferts in green and LINERs in red. In each panel, sources to the left of the dashed line ($log~R=2.4$) are radio quiet, while sources to the right are radio loud (Panessa et al. 2007). The overall Effelsberg sample is shown in grey.} 
 \label{R}
 \end{figure*}
 \begin{table*}
  \caption{\label{R_tab}Statistics on the radio loudness of the sample.}
  \centering
    \begin{tabular}{|cc|cc|cc|cc|cc|}
    \hline
     \multicolumn{2}{|c}{SFGs}&\multicolumn{2}{c}{Composites}&\multicolumn{2}{c}{Seyferts}&\multicolumn{2}{c|}{LINERs}\\
     \multicolumn{2}{|c}{6}&\multicolumn{2}{c}{19}&\multicolumn{2}{c}{42}&\multicolumn{2}{c|}{52}\\
    \hline
    \hline
     R mean&R median&R mean&R median&R mean&R median&R mean&R median\\
      2.66& 2.61& 2.61& 2.58& 2.74& 2.72& 2.83& 2.70\\
    \hline
     Quiet&Loud&Quiet&Loud&Quiet&Loud&Quiet&Loud\\
      2& 4& 4& 15& 8& 34& 7& 45\\
    \hline
    \end{tabular}
   \tablefoot{Row.1: Number of galaxies per spectral class as inferred from the [NII-based diagram]. Row.2: Mean and median values of $R$ (radio loudness). Row.3: Number of radio-quiet and radio-loud objects per spectral class.}
 \end{table*}

\subsection{Data reduction}
The data reduction, from raw telescope data to calibrated spectra,
was done in the standard manner \citep[see e.g.][]{2008A&A...490.1019F}
and included the following steps: (i) baseline subtraction and Gauss fitting 
of (sub-) scans and the necessary steps of data quality control 
(on sub-scan level) including flagging of discrepant or otherwise 
bad sub-scans/scans, (ii) correction of small residual pointing 
offsets (typically $\le\,5-10$\,''), (iii) atmospheric opacity 
correction using the $T_{\mathrm{sys}}$-measurements obtained with 
each scan, (iv) accounting for remaining systematic gain-elevation 
effects and finally (v) flux density calibration using the frequent 
primary calibrator measurements 
\citep[][]{1977A&A....61...99B,1994A&A...284..331O,2008ApJ...681.1296Z}. 
The overall measurement uncertainties are less than a few percent.

\begin{figure}[!ht] 
  \centering
  \includegraphics[width=9cm]{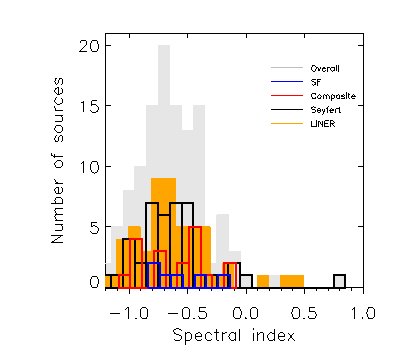}
  \caption{Two-point spectral index distribution ($\alpha_{[4.85-10.45]}$). Colors indicate the spectral classes according to the [NII]-based diagnostic diagram. The overall Effelsberg sample is shown in grey.}
 \label{alpha}
 \end{figure}

\section{Analysis and Results}
\subsection {Radio loudness}
Galaxies can be classified into two families, radio quiet and radio loud, depending on their monochromatic luminosity at radio ($L_R$) and optical ($L_O$) wavelengths. In this work, radio loudness ($R$) is defined as in \citet{Panessa2007}, as $L_{4.85}$/$L_{O}$, where $L_{4.85}$ is the integrated radio flux density at $4.85$ GHz. For $R \leq 10^{2.4}$, sources are considered to be radio quiet, while for $R > 10^{2.4}$, galaxies are considered to be radio loud. Note that this classification is to some degree arbitrary, as there is a continuum of radio loudness values, with no obvious bimodality in this distribution. Radio loudness suffers from selection effects. As radio galaxies are selected to have high radio luminosities, this results in large values of $R$ and the lack of many low-power galaxies and AGNs \citep{Sadler2002}. This is the case for our sample, where the lower limit in radio flux density ($F_{\rm 1.4} \geq 100$ mJy) selects strong radio emitters, leading to a quite severe selection bias. In 
the optical regime, radiation from star-forming galaxies could be suppressed by dust, which would result in high radio-optical flux ratios \citep{Afonso2005}. These sources are, however, mostly revealed as submillijansky and microjansky radio sources and are placed at high redshifts \citep[$z \gtrsim 1$, see][]{Barger2000}, and are excluded from our sample.
In addition, the observed optical emission of type 2 sources suffers from dust obscuration/extinction leading to an overestimation of radio loudness. The contribution of different components/processes to the radio emission (starburst activity, AGN) is a further complication. We refer to e.g. \citet{Sikora2007} for a detailed
discussion on radio loudness related issues.
Another problem arises when computing the optical luminosity. A correct computation of the $R$ parameter would imply a method to subtract the contribution of the host to the overall optical emission, e.g. via galaxy image decomposition \citep{Ho2014}. This technique, although powerful, can be used on high-resolution images only, and at best for local sources. When the host-light subtraction is not achievable, cetral optical luminosity is the next good option \citep{Ho2001}. Unfortunately, the Effelsberg sample has been selected to avoid aperture effects. Thus, the $3\arcsec$ SDSS fibers collect light from the galaxy nuclear regions as well as from the bulge, and we cannot retrieve central luminosities.
All this evidence suggests that radio loudness varies strongly for different galaxy populations, and it is arduous to estimate $R$ for a mixed sample of radio-bright star-forming galaxies and AGNs placed at intermediate redshifts. We nevertheless calculate the radio loudness seeking for broad trends in our sample.

We calculate $R$ by using the logarithmic ratio between the $4.85$-GHz flux density (measured with Effelsberg) and $R$-band observed flux density in Jy (calculated from SDSS DR7). We find that $98$ ($82.3$\%) of our radio emitters are radio loud, and $21$ ($17.6$\%) are radio quiet. Radio loudness for each spectral class is shown in Fig. \ref{R}, and the statistics are listed in Tab. \ref{R_tab}. LINERs are the most radio loud objects of the sample (mean $R= 10^{2.83}$, median $R= 10^{2.70}$) and their $R$ reaches values of $R \sim 10^{4.70}$. Seyferts and LINERs show a gaussian distribution, with peaks at $R \sim 10^{2.7}$.There is no obvious bimodality in $R$ as seen in the overall Effelsberg sample (Fig. \ref{R}). The dividing line, though, corresponds well to the location of the peak in the observed radio loudness distribution.

\begin{figure*}[!ht] 
  \centering
  \includegraphics[width=9cm]{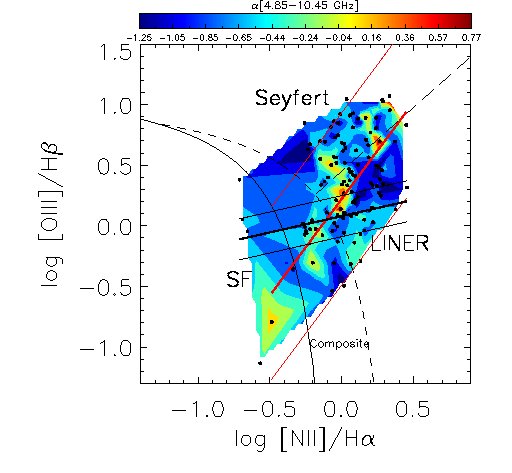}
  \caption{Two-point spectral index distribution of the Effelsberg sample represented in the [NII]-based diagnostic diagram. The color gradient indicates the $\alpha_{[4.85-10.45]}$ values. Black dots correspond to sources positions in the diagram. Red thick lines are regression curves of the $15$\% most flat- and inverted-spectrum sources, while black thick lines are regression curves of the steep-spectrum sources. $\pm 1 \sigma$ of regression curves are represented by the outer red and black lines.}
 \label{alpha_BPT}
 \end{figure*}

\begin{figure*} 
  \centering
  \includegraphics[width=18.5cm]{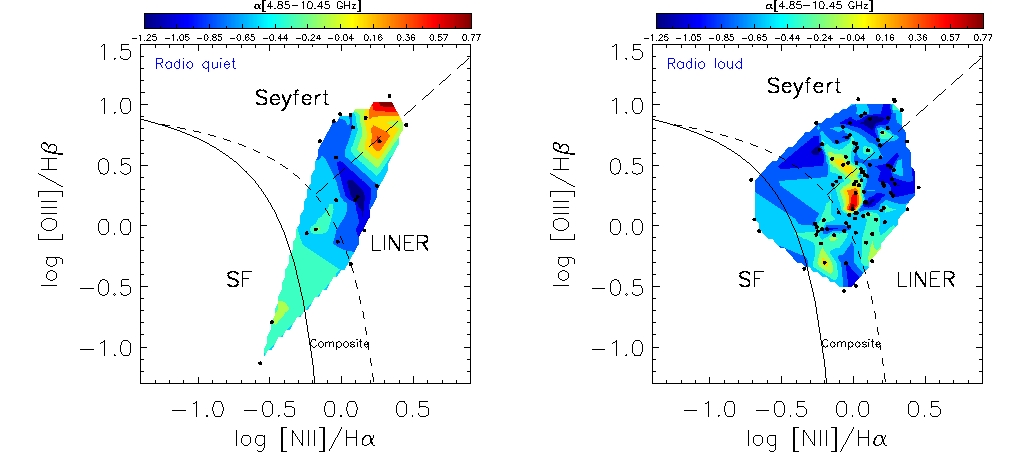}
  \caption{As in Fig. \ref{alpha_BPT}, but divided into radio quiet (left) and radio loud (right) galaxies. The radio loudness limit ($R=2.4$) follows \citet{Panessa2007}.}
 \label{R_BPT}
 \end{figure*}

\subsection{Spectral indices}
We calculate the two-point spectral index from our simultaneous Effelsberg measurements at $4.85$ and $10.45$ GHz ($\alpha_{[4.85-10.45]}$). The spectral index $\alpha$ is defined as $F(\theta) \propto \nu^{\alpha}$ and is equal to:
\begin{equation}
 \alpha = \frac{\log \frac{F_{4.85}}{F_{10.45}}}{\log \frac{10.45}{4.85}}
\end{equation}
where $F_{4.85}$ and $F_{10.45}$ are the observed flux densities at $4.85$ and $10.45$ GHz.
Spectra are here considered as \textit{steep} when $\alpha \leq -0.5$, \textit{flat} when $-0.5 < \alpha \leq 0$, and \textit{inverted} when $\alpha > 0$.
The mean (median) error on the spectral index is $\sigma_{\alpha}=0.22$ ($\sigma_{\alpha}=0.14$), estimated from the standard propagation of errors:
\begin{equation}
 \sigma_{\alpha}= \frac{1}{log(10.45/4.85)} \times \sqrt{\left(\frac{\sigma_{F_{4.85}}}{F_{4.85}}\right)^2 + \left(\frac{\sigma_{F_{10.45}}}{F_{10.45}}\right)^2}
\end{equation}
where $\sigma_{F_{4.85}}$, $\sigma_{F_{10.45}}$ are the respective flux density errors.

For comparison, we also calculate the two-point spectral index using the $1.4$ GHz (FIRST) and $4.85$ GHz (Effelsberg) non-simultaneous data ($\alpha_{[1.4-4.85]}$).
We then calculate the spectral curvature, defined as $\alpha_{[4.85-10.45]} - \alpha_{[1.4-4.85]}$, following \citet{Gregorini1984}. Curvature classes are: \textit{concave} ($C\geq 0.01$), \textit{flat} ($-0.1<C<0.1$), and \textit{convex} (or \textit{humped}, $C\leq -0.01$) spectra.

Fig.\ref{alpha} shows the two-point spectral index distributions as derived from Effelsberg data at $4.85$ and $10.45$ GHz for star-forming, composite, Seyfert and LINER galaxies. Seyferts and LINERs span a wider $\alpha$ range than star-forming galaxies and composites. The range is $-1.25< \alpha < 0.85$ for Seyferts, and $-1.25< \alpha < 0.5$ for LINERs, while it is $-1.1< \alpha < -0.1$ for composite and $-0.85< \alpha < -0.15$ for star-forming galaxies. Mean (median) $\alpha$ values per spectral class are: $-0.59$ ($-0.61$) for star-forming, $-0.61$ ($-0.54$)  for composite, $-0.62$ ($-0.64$) for Seyfert, and $-0.62$ ($-0.67$) for LINER galaxies. We do not find significant differences between the galaxy classes.
The distribution of the overall sample shows a peak at $\alpha \sim -0.7$, and a second tentative one at $\alpha \sim -0.2$. As suggested by a previous study based on $1.4$-GHz observations of $345$ objects from NRAO-MPI $4.85$-GHz surveys of extragalactic sources \citep{Witzel1979}, the distribution of the two-point spectral indices at these frequencies exhibits a double peak. \citet{Witzel1979} found that galaxies are mainly responsible for the peak at $\alpha \sim -0.8$, while quasars form the peak at $\alpha \sim 0$. We do not see this galaxy/quasar division as our redshift limits and exclusion of broad emission line systems ensures we have very few quasar-like objects in our sample. 

Fig. \ref{alpha_BPT} shows the two-point spectral index distribution in the [NII]-based diagnostic diagram. Steep spectra are shown as blue, flat spectra green, and inverted spectra yellow-red.
The red line represents the regression curve 
of the $15$\% most flat- and inverted-spectrum sources ($18$ galaxies). The parallel red thinner lines are the $+1 \sigma$ and $-1 \sigma$ lines. 
The black line is the regression curve 
of all steep-spectrum sources ($\alpha_{[4.85-10.45]}\leq -0.5$, $82$ galaxies in total, $69$\% of the sample). Parallel black thinner lines are $\pm 1 \sigma$.
Flat and inverted sources in our $119$-galaxy sample are mainly clustered along a diagonal line. The line extends from the end of the star-forming sequence to the Seyfert region. These trends suggest a possible spectral ``flattening sequence'', which crosses the composite region of the diagram, and roughly follows the division line between Seyferts and LINERs. An extra region of spectral flattening is located at the bottom of the LINER region, close to the composite-LINER division line. For a sample which also includes Effelsberg measurements of sources with S/N$<3$ ($31$ additional galaxies, for a total of $150$ objects), the trend is similar. The lack of S/N cut recovers, however, mostly LINERs with steep spectral index.

\begin{figure}[!ht] 
  \centering
  \includegraphics[width=9.2cm]{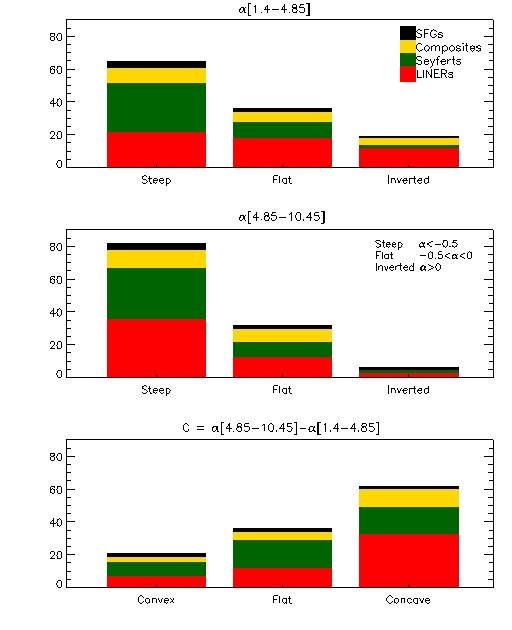}
  \caption{Spectral indices and spectral curvature for each spectral class. The top panel shows the cumulative histograms of the spectral index $\alpha_{[1.4-4.85]}$, while the middle panel shows $\alpha_{[4.85-10.45]}$. Spectra are classified as steep, flat, or inverted. The bottom panel shows the spectral curvature ($C$), defined as the difference between $\alpha_{[4.85-10.45]}$ (higher frequency) and $\alpha_{[1.4-4.85]}$ (lower frequency).}
 \label{curvature}
 \end{figure}

In Fig. \ref{R_BPT} we study whether this trend is persistent in the case of separated radio quiet ($21$) and radio loud ($98$) objects. 
We conclude that no clear trend between radio-quiet and radio-loud objects is visible, and the spectral index flattening sequence mostly results from the combination of radio quiet high-metallicity star-forming galaxies and [O\,{\sc{iii}}] emitters (Seyferts/LINERs), and radio-loud LINERs and composites. It is nonetheless tantalising to note that the most extreme inverted radio spectra correspond to the most extreme optical line ratios for radio {\em quiet} systems, while for radio {\em loud} they seem to favour optical line diagnostics close to the \citet{Kewley2001} dividing line.


Fig. \ref{curvature} shows cumulative histograms of the spectral indices ($\alpha_{[1.4-4.85]}$ and $\alpha_{[4.85-10.45]}$) and curvature, calculated for each galaxy class. The top panel shows that $65$ out of $119$ Effelsberg sources have steep spectra between $1.4$ and $4.85$ GHz. Seyfert galaxies dominate the distribution ($30$ sources), though they are not the most abundant spectral class in our sample. The other steep-spectrum sources are: $4$ star-forming, $9$ composite, and $22$ LINER galaxies. On the other hand, LINERs show the highest number of flat and inverted spectra at these frequencies ($18$ and $12$, respectively). At the higher Effelsberg frequencies ($4.85$-$10.45$ GHz, middle panel), the number of steep-spectrum sources slightly increases ($82$ instead of $65$), especially in case of LINERs (which are now $36$). The number of star-forming, composite, and Seyfert galaxies remains almost constant compared to the $1.4$-$4.85$ GHz frequency range ($4$, $11$, and $31$, respectively). \
The bottom panel shows the spectral curvature trends. About $50\%$ of sources display a negative curvature (convex shape), with high fluxes at the intermediate frequency ($4.85$ GHz). Seyferts dominate the ``concave'' (positive curvature) and ``flat'' (zero curvature) distributions ($9$ out of $21$ and $17$ out of $36$ sources, respectively), while LINERs are the class of objects showing the highest number of ``convex'' spectra ($33$ over $62$ total sources with negative curvature), peaking at $\sim 5$ GHz.  


GHz-peaked spectra are sometimes an indication of extremely compact sources \citep[GHz peaked-spectrum, or compact steep-spectrum sources,][]{O'Dea1998}, whose nucleus is bright at high radio frequencies (up to $\sim 5$ GHz). A concave shape often results, in contrast, from the steep-spectrum contribution of extended structures (lobes) at $\nu \lesssim 5$ GHz, and the flat/inverted contribution of the nucleus. The latter makes the spectra rise at $\nu \gtrsim 5$ GHz. 
However, it must be mentioned that with our Effelsberg data ($4.85-10.45$ GHz) we are mainly looking at radio emission from the central region, or at a contamination of quiescent jet emission and activity in the nucleus. This is because the lobes, when present, mostly emit at lower frequencies, due to their steep spectra and the aging of the synchrotron radiation. Therefore, we assume that the flattening of the spectral indices come almost exclusively from nuclear activity. Following the relation between turnover frequency, source size, turnover frequency and peak flux density, as well as magnetic field strength \citep{KellermannPauliny-Toth1981} our results imply that the high frequency emission of Seyferts along the division line between Seyferts and LINERs in the [NII]-based diagnostic diagram is dominated by compact radio emission from central regions (flat spectra), whereas the emission of LINERs is dominated by extended emission from larger source components (steep spectra). Those could be due to 
arcsecond sized components that may be associated with nuclear jets, hot spots in radio lobes, or very compact wind/shock regions that are bright in the radio domain.
\begin{figure}[!ht] 
  \centering
  \includegraphics[width=9.5cm]{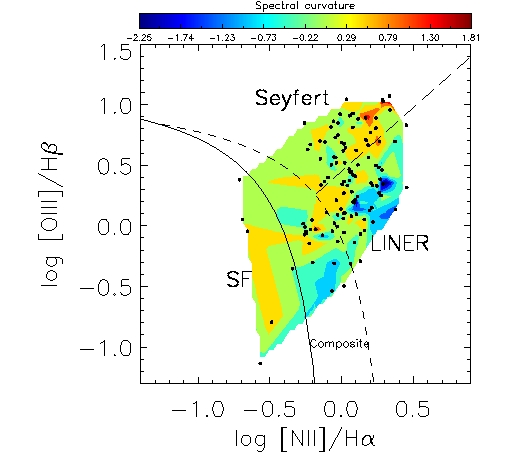}
  \caption{Spectral curvature distribution in the [NII]-based diagnostic diagram. The color indicates the spectral curvature values. Black dots correspond to sources positions. $C$ is defined as $\alpha_{[4.85-10.45]} - \alpha_{[1.4-4.85]}$.}
 \label{curvature_BPT}
 \end{figure}

Spectral information includes the effects of variability, i.e. changes in the spectral slope between $1.4$-$4.85$ GHz and $4.85$-$10.45$ GHz. This might be due to flux increase at the highest frequencies of our spectral range, because of the activity in the compact nucleus on time scales of days to years. FIRST data were collected about $20$ years ago, so a change in the spectral index may be associated with nuclear variability. However, for the non-blazar like objects considered in this study, variability is in general not high in the spectral interval $1.4-10.45$ GHz \citep{Eckart1989}. This means that we can safely make assumptions on the shape of the spectra, without taking variability into account.


Fig. \ref {curvature_BPT} illustrates the color-coded spectral curvature distribution in the [NII]-based diagram. The diagram shows a clearer difference between Seyferts and LINERs with respect to the spectral index distribution in Fig. \ref{alpha_BPT}. This is because Seyferts tend to have a zero or positive curvature, indicating flat or concave-shaped spectra, while LINERs show more convex spectra.


\begin{figure}[!ht] 
  \centering
  \includegraphics[width=9cm]{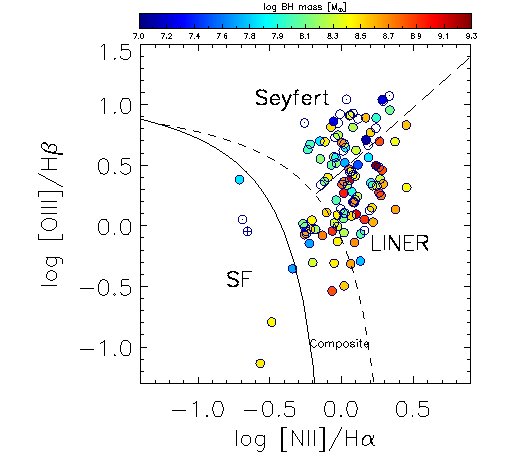}
  \caption{Black hole masses distribution in the [NII]-based diagram. The color bar indicates $M_{\rm BH}$ in solar masses. White circles indicates sources where the SDSS measurement of the stellar velocity dispersion is not accurate (flags $-3$ and $-50$). The crossed circle indicates again a non reliable measurement, not flagged in the SDSS catalog.}
 \label{BH_mass_BPT}
 \end{figure}

\subsection{Black hole mass}
Following \citet{Gueltekin2009}, we calculate the black hole masses for each source in the Effelsberg sample from the scaling relation with stellar velocity dispersion ($\sigma$), which are provided in the SDSS DR7 and measured from the galaxy absorption lines:
\begin{equation}
 M_{\rm BH} = 8.12+4.24\times \log\frac{\sigma}{200}
\end{equation}
Fig. \ref{BH_mass_BPT} shows the $M_{\rm BH}$ distribution in the [NII]-based diagram. Black-hole masses range from $10^6$ to $10^{9.3} M_{\odot}$. The objects with the highest black hole masses are mostly located in the composite and LINER regions of the diagram. Mean (median) logarithmic black-hole masses are : $8.1$ ($8.5$) for star-forming, $8.4$ ($8.5$) for composite, $8.1$ ($8.2$) for Seyfert, and $8.5$ ($8.5$) $M_{\odot}$ for LINER galaxies. Assuming the $M_{\rm BH}-M_{\rm bulge}$ relation (see Sect. 5.1), this trend points to LINERs being the oldest and most evolved spectral class in the diagram. We need to point out, however, that these differences in $M_{\rm BH}$ among the four galaxy spectral classes may not be significant, since for each object the $M_{\rm BH}$ calculation relies on the accuracy of the stellar velocity dispersion measurements.

For a few objects, indicated with white circles in Fig. \ref{BH_mass_BPT}, the determination of the black hole mass is not reliable. Reliability is estimated according to SDSS error flags on the measurement of $\sigma$. We notice that the latter objects are not isolated cases of stellar velocity dispersion mismeasurements, but they occupy precise regions of the diagnostic diagrams of the cross-matched SDSS-FIRST parent sample, namely the upper part of the Seyfert region (see Fig. \ref{BH_mass_BPT_parent}, right panel). 
We checked the SDSS morphology, and based on it we speculate that the Seyferts with very high [O\,{\sc{iii}}]/H$\beta$ could be identified with the newly discovered population of ``green beans'' galaxies \citep{Schirmer2013}. This class of objects is indicated with red triangles in Fig.\ref{BH_mass_BPT_parent}. Those which are Seyferts show especially high log [O\,{\sc{iii}}]/H$\beta$ ratios ($>1$), appear green in the SDSS images, and often show signs of perturbed morphology and post-merger activity. Those perturbations the green beans undergo could be an explanation of the stellar velocity dispersion mismeasurements.

A second class of objects showing velocity dispersion mismeasurements, this time not flagged in the SDSS DR7, is represented by only two star-forming sources in the Effelsberg sample (indicated with a white circle and a crossed white circle in Fig. \ref{BH_mass_BPT}) and by a larger SDSS-FIRST parent population (blue squares in Fig. \ref{BH_mass_BPT_parent}). 
Those galaxies may be associated to the ``green peas'' \citep{Cardamone2009}. Green peas are starburst galaxies and are very compact. They also show signs of merger activity. According to \citet{Cardamone2009}, the green peas might be the downscaled version of high-redshift massive merging galaxies, observed in the local universe due to the mass downsizing (less massive galaxies evolve in a longer interval of time). 

In Fig. \ref{BH_mass_BPT} we see an apparent trend to higher-mass black holes for LINER systems compared to Seyferts, increasing initially between star-forming galaxies to Seyferts. Aknowledging the small number statistics, and the presence of unreliable measurements in the sample, this trend still supports a possible evolution from star-forming to LINER galaxies. The black-hole mass distribution of the parent sample (Fig. \ref{BH_mass_BPT_parent}), thanks to a much bigger sample size, points to a clear transition from low-metallicity star-forming galaxies to composites and LINERs, where the latter are found to be the galaxies harboring the most massive black holes in the sample.
 \begin{figure}
  \centering
  \includegraphics[width=8cm]{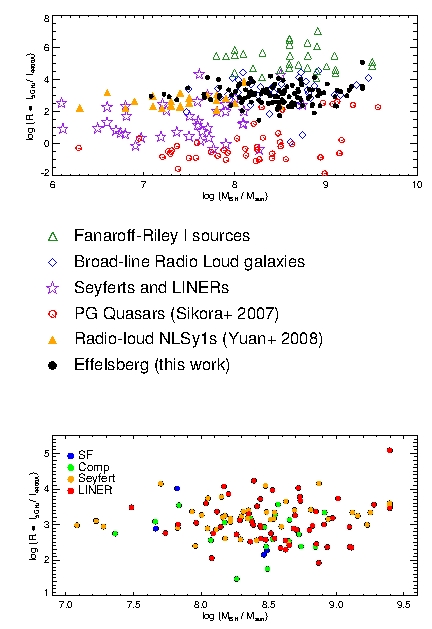}
  \caption{Black hole mass versus radio loudness for some selected galaxy samples (top panel) and for the Effelsberg sample, divided by spectral classes (bottom panel).}
 \label{M_BH_vs_R}
 \end{figure}

The upper panel of Fig. \ref{M_BH_vs_R} shows the position of the Effelsberg objects in the black-hole mass-radio loudness ($M_{\rm BH}$-$R$) plane. A set of different galaxy classes from \citet{Sikora2007} are represented in different colors and symbols: FR I, broad-line radio loud galaxies, Seyferts and LINERs, and Palomar-Green (PG) quasars. Radio-loud Narrow-Line Seyfert 1s (NLSy1s) are from \citet{Yuan2008}. Our sample (black dots) is mostly consistent with broad-line radio loud galaxies and partly with radio loud NLSy1s, but not with Seyferts and LINERs. This finding is consistent with our previous discussion on the radio flux lower cut of the Effelsberg sample, which mostly selects powerful radio-loud objects with high $M_{\rm BH}$. 
The lower panel shows our Effelsberg sample divided by spectral class, according to the [NII]-based diagnostic diagram. The four spectral classes are almost equally spread in the $M_{\rm BH}$-$R$ plane. 

\subsection {Radio and Optical morphologies}
On one hand, a host's optical morphology can reveal signs of galaxy-galaxy interactions, and information on galaxy stellar and gas contents. Spectroscopic surveys find a link between emission type (as classified in the optical emission-line diagnostic diagrams) and Hubble type \citep{Ho1997}. HII emission tends to be in late type spirals and ellipticals, including low-luminosity galaxies. LINERs and Seyferts tend to inhabit early-type spirals and to avoid low-luminosity galaxies instead, suggesting their presence in more evolved hosts \citep{Moles1995,Ho1997}. On the other hand, radio morphology can give important hints on the evolutionary stage of a galaxy. For instance, lobe extensions and shapes can be used to measure the AGN age \citep{Shulevski2012}. The presence of jets in radio images helps to constrain cases where AGN feedback can regulate star formation.

In this framework, we have visually analyzed the FIRST radio images of the Effelsberg sample. About $70$\% of the objects are unresolved at FIRST resolutions ($\sim 5$''), and they appear as point sources in the images. The resolved sources show extended radio lobes, and are almost exclusively classified as LINERs. In particular, we find $21$ galaxies ($18$\% of the sample) showing lobes or jets in FIRST images, and $12$ of them are LINERs according to the [NII]-based emission-line diagnostic diagram (see Tab. \ref{all_info}). This is in agreement with our hypothesis that LINERs might be dominated by lobe emission in the radio domain, while Seyferts are core dominated. Other sources with extended radio emission are classified as Seyferts ($6$) and composites ($3$). $15$ sources show less clear signs of extended emission (e.g. elongated structures, not clearly identifiable with lobes). These are again mostly LINERs ($11$). Finally, $6$ galaxies show multiple point sources in the radio images.
 This could be either due to the 
presence of bright dense star-forming regions (e.g. in case of starburst galaxies) or the presence of an unrelated source of radio emission.

\begin{figure}
  \centering
  \includegraphics[width=9cm]{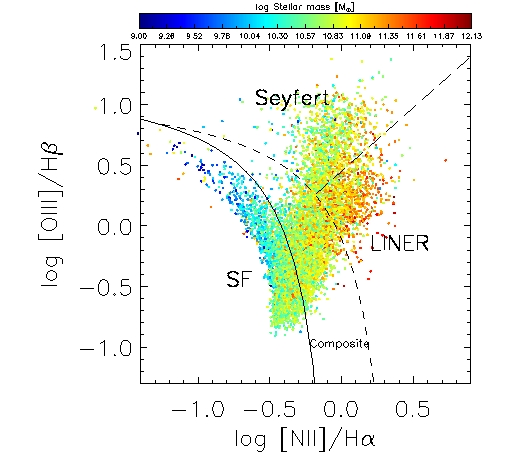}
  \caption{SDSS-FIRST stellar mass distribution in the [NII]-based diagnostic diagram. The color bar indicates the stellar mass values from SDSS measurements, in solar units.}
 \label{stellar_mass_BPT}
 \end{figure} 
   
SDSS optical images of our sample are the only ones available for our objects, given their intermediate redshift. Images do not show peculiar morphologies, except for a few post-merger cases ($3$) and galaxies with signs of a possible interaction with neighboring galaxies ($5$). A detailed analysis of the optical morphology is beyond the scope of the current work. However, we notice that galaxies with extended features such as lobes and jets show lenticular (S0) and elliptical (E) morphology. Often, it is not possible to distinguish between the two. SDSS images of our sample, separated by spectral class, can be found in the Appendix (Fig. \ref{sfgs}-\ref{liners}).

The merging and interaction signatures in the SDSS images of the parent sample are present in galaxies belonging to the upper part of the star-forming sequence (low-metallicity star-forming galaxies). Unfortunately, low-metallicity starbursts are under-represented due to the $1.4$-GHz flux lower limit, and only two candidate post-merger galaxies are found in the Effelsberg sample (Fig. \ref{sfgs}, panel 2 and 5). A different sample (i.e. with a lower radio flux limit) might be able to recover more low-metallicity starbursts and improve the poor statistics, revealing that interactions are a common feature in this class of young objects. The $1.4$-GHz flux lower limit still selects many Seyfert galaxies. Seyferts with perturbed morphologies might be more abundant in the Effelsberg sample, where we count $19$ galaxies with stellar velocity dispersion mismeasurements.
In particular, we see post-merger signs in the Seyferts with high [O\,{\sc{iii}}]/H$\beta$, a characteristic that was already noted by \citet{Schirmer2013} and may have important implications for the development of a galaxy evolution scheme accross the diagnostic diagrams. Their SDSS colors are also redder than those of galaxies placed in the star-forming region.

\begin{figure*}
  \centering
  \includegraphics[width=17cm]{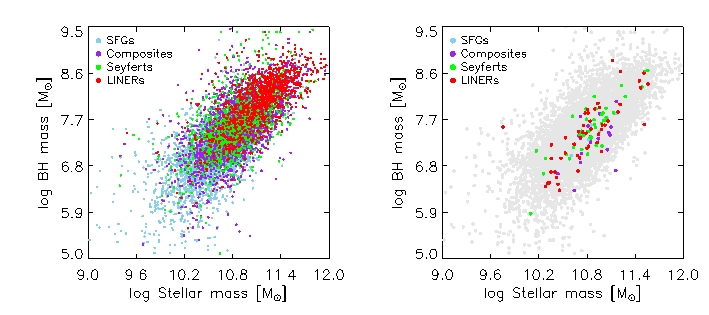}
  \caption{Trend between stellar mass and BH mass of the parent sample (left panel) and the Effelsberg sample (right panel). Colors indicate the spectral classes. In the right panel, the grey circles indicate the parent sample.}
 \label{correlation}
 \end{figure*}

\section {Discussion}
\subsection{$M_{\rm BH}-M_{\rm bulge}$ correlation}
Studies on host-dominated AGNs suggest that AGNs are more likely to be hosted in massive galaxies \citep{Kauffmann2003b,Haggard2010,Vitale2013}, though \citet{Aird2012} claim that the trend is due to the Eddington-ratio distribution: a higher fraction of AGNs are detected in massive hosts as they are intrinsically more luminous. \citet{Vitale2013} show that up to intermediate redshifts ($z \sim 1$) AGN hosts have stellar masses of $M_* \gtrsim  10^{10.2} M_{\odot}$, and that mass increases along the left branch of the distribution in the [NII]-based diagram \citep[see also][]{Kauffmann2003b,Stasinska2006}. Low-metallicity star-forming galaxies have, therefore, lower stellar masses than metal-rich star-forming galaxies. Star-forming galaxies have, in general, lower stellar masses than composites and AGNs. This means that, by selecting mailnly metal-rich star-forming galaxies, composites, and AGNs from the optical-radio domain \citep{Vitale2012}, we automatically exclude galaxies with less massive hosts from 
the sample.

Fig. \ref{stellar_mass_BPT} shows the SDSS-FIRST stellar mass distribution in the [NII]-based diagram. Stellar masses are provided in the MPA-JHU DR7 of spectrum measurements\footnote{http://www.mpa-garching.mpg.de/SDSS/DR7/}, as mentioned in Sect. 2.1. Metal-poor star-forming galaxies exhibit the lowest stellar mass values ($M_* \lesssim 10^{10}$ $M_{\odot}$), while LINERs the highest ($M_* \gtrsim 10^{11}$ $M_{\odot}$)\footnote{Data on galaxy stellar masses from the MPA-JHU emission line analysis of SDSS DR7}. Average black-hole masses are low for star-forming galaxies ($M_{\rm BH} \lesssim 10^{7}$ $M_{\odot}$), increase for composites and Seyfers and are one order of magnitude higher than in SFGs for LINERs. This result is in agreement with LINERs mostly sitting in large ellipticals \citep{Ho1997}, and with the $M_{\rm BH}$ trend in Fig. \ref{BH_mass_BPT}. The trend is also in agreement with the studies of the correlation between small scales properties of the central black hole and large scale properties 
of galaxies in the local universe \citep{Ferrarese2000,Gebhardt2000,Marconi2003,Gueltekin2009,Kormendy2013}, though the estimation of black-hole masses relies of different methods and it is still affected by large errors.

The question arises whether the stellar mass is always related to the mass of the black hole. In Fig. \ref{correlation} we show the trend between average stellar SDSS masses and BH masses of the parent sample (left panel) and the Effelsberg sample (right panel). The star-forming galaxies appear to have the lowest stellar and BH masses. The distribution shows a progressive increase in both values for composites, Seyferts and LINERs. The Effelsberg sample, due to the smaller sample size, does not show the same clear trend.

The trend between stellar and BH mass, here shown for each spectral class of the SDSS-FIRST and Effelsberg samples, could be interpreted as the more massive and more high metallicity galaxies being the ones where the AGN can be ``switched on'' (Seyferts and LINERs). Stellar mass build up and BH accretion are connected and it is possible that AGN-feedback manifests itself in massive and more evolved hosts, with $M_{\rm BH} \gtrsim 10^7$ $M_{\odot}$. This consideration does not include Narrow-Line Seyfert 1 galaxies, which are anyhow rare objects with low $M_{\rm BH}$.

\subsection {An evolutionary sequence in the [NII]-based diagram?}
\begin{figure}
  \centering
  \includegraphics[width=9.5cm]{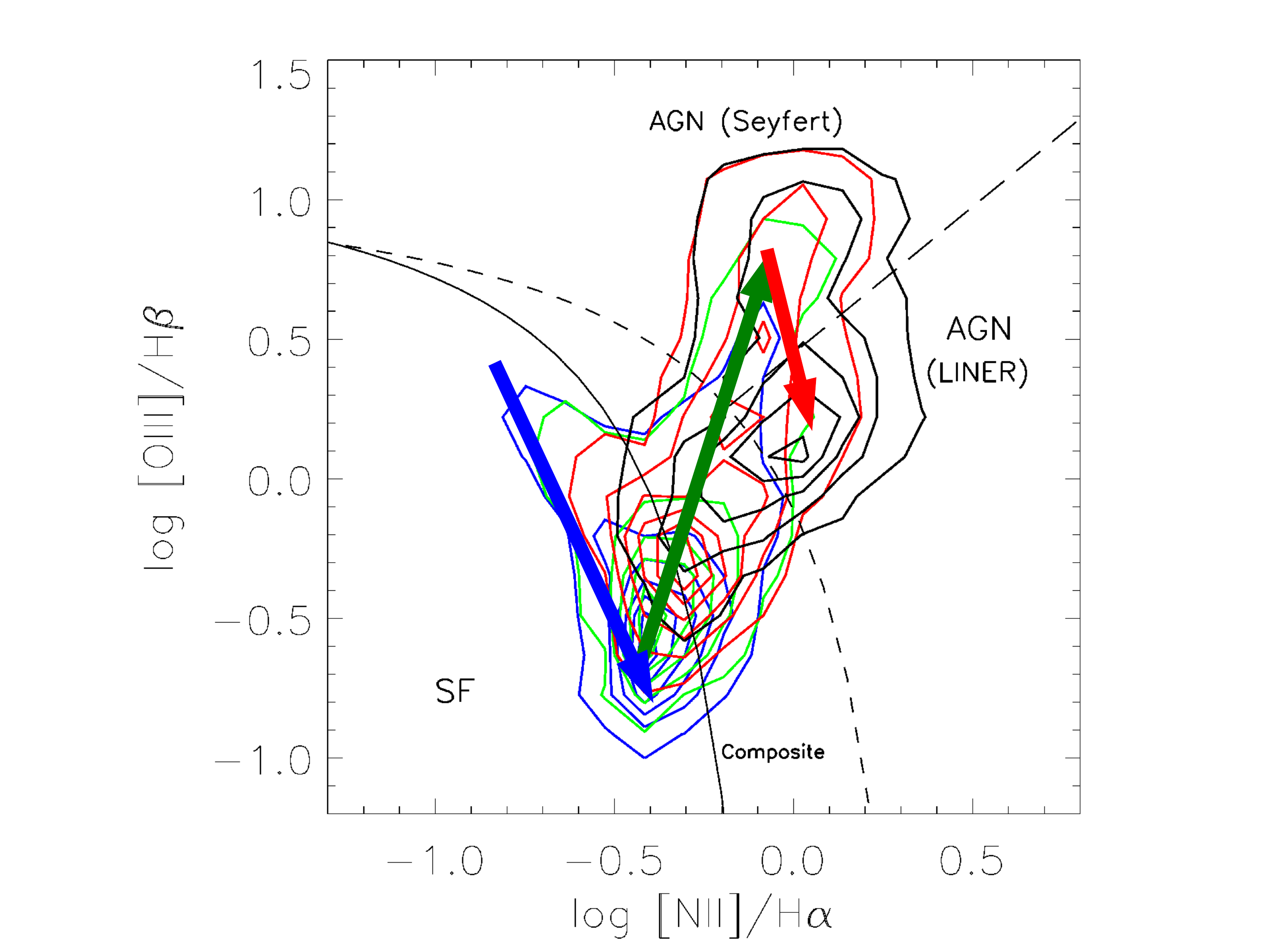}
  \caption{Sketch of galaxy evolution across the [NII]-based diagnostic diagram. Colour contours represent sub-samples of the parent sample with increasing values (blue, green, red and black) of the ratio between radio luminosity and luminosity of the H$\alpha$ line as in  \citet{Vitale2012}. The arrows represent the trend of possible galaxy evolution from star-forming galaxies to Seyferts and LINERs.}
 \label{sketch}
 \end{figure}
 
An important next step would be to set a simple and universally valid galaxy evolutionary scenario that explains how blue star-forming galaxies turn into red passively-evolving systems. Some models take into account this kind of transition and the necessity of a mechanism to truncate star formation \citep[eg.][]{Hopkins2006a}, but there is still the need of observational confirmations. The current standard models of galaxy evolution \citep{Hopkins2008b,Hopkins2008a} rely on major mergers and/or interactions to trigger both star formation and AGN activity in two merging late-type galaxies. Then the system relaxes and forms an early-type galaxy hosting a radio AGN. 
We are interested in testing a scenario where the end of the star-forming sequence represents a turnover point in galaxy evolution, i.e. the start of the quasar phase, which leads to the colour and morphological transformation of the hosts (transition from the active blue cloud to the ``dead'' red sequence).

The Effelsberg observations help us to probe galaxy spectral evolution in radio emitters that are classified as star-forming, composite, Seyfert and LINER galaxies in the optical emission-line diagnostic diagrams, by looking at their combined optical-radio properties. In particular, we see indications of a flattening of the spectral index along the composite-AGN sequence of the diagnostic diagrams. A flattening of the spectral index is expected due to the presence of compact nuclei/jet emission, that could indicate the presence of AGN activity as mechanism that shuts star formation down. A similar sequence was first noticed in \citet{Vitale2012,Vitale2013}. \citet{Vitale2012} have shown that the radio luminosity at $1.4$ GHz progressively increases from star-forming galaxies to Seyferts and LINERs. The H$\alpha$ luminosity - a tracer of star-formation activity - decreases in the same direction. This indicates that strong optical emission is a characteristic sign of recent star-formation, 
thus of the youth of a galaxy, while radio emission appears as a ``later'' feature, in massive and red ellipticals. \citet{Vitale2013} found that galaxy stellar masses, and age and metallicity of the stellar populations, increase progressively from star-forming galaxies and composites, to AGNs. These findings strengthen the hyphothesis of a galaxy evolutionary sequence based on the analysis of optical and radio properties of normal galaxies and AGN hosts.

In additional support to this scenario, other studies have already been conducted on separate optical and radio samples. Of special interest is the study of high- and low- excitation systems \citep[see e.g.][]{Wierzbowska2011}. It has been suggested that the two major radio AGN populations - the powerful ($L_{1.4~GHz}>10^{25}$ W/Hz) high-excitation, and the weak ($L_{1.4~GHz}<10^{25}$) low-excitation radio AGNs - represent an earlier and later stage of massive galaxy build-up \citep{Hardcastle2006,Smolcic2009}. As support for this hyphothesis, a clear dichotomy is found between the properties of low-excitation (LEGs) and high-excitation (HEGs) galaxies. In the radio, FR I are mostly LEGs, while some FR II are HEGs. HEGs are hypothesised to evolve into LEGs due to changes in the black-hole accretion rate, maybe after accreting material is no longer available. The hosts of LEGs have the highest stellar masses, reddest optical colors, and highest black holes masses, but accrete at low rates. 
On the other hand, the high-excitation radio AGNs have lower stellar masses, bluer optical colors (consistent with the green valley), lower mass black holes that accrete at high rates \citep{Smolcic2009}.

A link could also exist between HEGs/LEGs and Seyferts/LINERs. Some optically-classified Seyferts appear to be mostly luminous FR II radio sources, while LINERs are more consistent with FR Is \citep{Buttiglione2010} and a ``later'' stage of radio evolution. FR II hosts were found to be bluer than hosts of FR Is and often showed signatures of mergers \citep{Heckman1986,Baldi2008,Ramos-Almeida2012}. Also, the host galaxies of FR Is were found to be more massive than the FR II hosts \citep{Owen1989,Govoni2000}. This might be due to the fact that also FRI and FRII have been hypothesized to be AGNs with different BH accretion rates \citep{Buttiglione2010}. These studies seem to point to an evolution from Seyferts to LINERs, possibly regulated by AGN-feedback. Therefore, LINER activity might be the ``smoking gun'' of the highly effective suppression mode (radio mode) of AGN feedback \citep{Schawinski2007}. Consequently the question we want to address is: are LINERs the last stage of galaxy evolution 
across the diagnostic diagrams?

In Fig. \ref{alpha_BPT}, the possible presence of a sequence of galaxies with flat or inverted spectra is a hint of increased nuclear activity along the same. The sequence could be explained as an increase in the strength of the ionising field of the galaxies, due to disk accretion. LINERs, which are mostly found in old hosts, do not show this feature and represent a different class of shock-ionised (rather than photo-ionised) objects. In LINERs, AGN activity could have led to negative feedback, where powerful jets and/or galactic superwinds blow most of the gas away and prevent star-formation.

As seen in Fig. \ref{alpha_BPT}, radio-AGN signature inferred from the spectra is also present in metal-rich star-forming galaxies. We find three possible scenarios to explain  flat-spectrum radio sources in the star-forming region of the diagnostic diagram:
\begin{itemize}
 \item The enhancement of star formation might be triggered by the AGN, which induces gas compression and acts as positive feedback.
 \item Sources host young AGNs, and the star formation in the galaxy has not yet been suppressed. This could be the case for strong and widespread star formation, where the timescale for it to cease could be several hundred Myr - potentially longer than the lifetime of the AGN. If AGNs are recurrent phenomena, it might be a periodic process for multiple episodes of the AGN that progressively slows down the star formation throughout the galaxy.
 \item The AGN is not strong enough to suppress the star formation entirely, or there is no link between the AGN and the star formation suppression. In this latter case, flat-spectrum radio galaxies in the star-forming region of the diagram would represent counter-examples to the hypothesis of AGN quenching star formation.
\end{itemize}
Finally, LINERS with $\alpha \sim -0.5$ (Fig. \ref{alpha_BPT}) might be older AGN with radio emission dominated by the echoes of old electron populations, as also evident from the radio images (Fig \ref{liners}). This fits with our nominal evolutionary sequence, and is consistent with the natural progression from that.

The importance of mergers at $z<1$ has been unclear, and assumed to be small, given the low observed merger rate \citep[e.g.][]{Lotz2008} before the discovery of Ultra-Luminous Infra-Red Galaxies (ULIRGs) and the ``green peas'' class of objects \citep{Cardamone2009}, which may represent the downscaled version of high-redshift merging galaxies. The low stellar-mass and low-metallicity galaxies merging at $z<1$ may well represent the beginning of an evolutionary sequence traced by the diagnostic diagrams at $z<0.4$ (Fig. \ref{sketch}). Along the star-forming sequence of the [NII]-based diagram and up to the composites-AGN region, the galaxy total stellar mass has been found to increase \citep{Vitale2013}, probably due to the presence of gas brought up by mergers/interactions. At the same time, the black hole is fed and its mass increases.

To summarize, while there are some studies aiming at explaining the most latest stages of galaxy evolution, observational proofs of the transition from star-forming to high-excitation systems are still missing, especially in the radio-optical regime. Our spectral index flattening sequence traces the increase of the ionising field from star-forming galaxies to AGNs, probably due to disk accretion. AGN activity may translate into feedback, which regulates star formation in the passage from composite to Seyfert and LINER galaxies.

\section{Conclusions}
We have conducted observations with the Effelsberg 100-m telescope to probe galaxy evolution in radio emitters that are classified as star-forming, composite, Seyfert and LINER galaxies in the optical emission-line diagnostic diagrams, by looking at their combined optical-radio properties. We have found indications for a flattening of the spectral index along the composite-AGN sequence of the diagnostic diagrams. This supports a scenario where the end of the star-forming sequence represents a turning point in galaxy evolution, i.e. the start of the quasar phase, which leads to the colour and morphological transformation of the hosts. The analysis of the spectral index distribution, combined with the information on the radio morphology (low resolution, from FIRST survey at $1.4$ GHz), have been presented to investigate the nature of the sources along the AGN branch, their active state, and the chance to spot an evolutionary sequence. Our conclusions are:
\begin{itemize}
 \item The objects in our sample show a possible sequence of flattening spectral indices that extend from the high-metallicity end of the star-forming sequence in the [NII]-based diagnostic diagram, to the Seyfert region, with the highest (most inverted) $\alpha_{[4.85-10.45]}$ values found among the AGNs. The sequence crosses the composite region and follows the Seyfert/LINER division line. It is consistent with a progressive hardening of the ionizing field of the galaxy, due to intense nuclear activity.
 \item The spectral curvature distribution shows the difference in the Seyfert and LINER spectra, with Seyferts having flatter and LINER more convex ($5$-GHz peaked) spectra. This suggests that the high frequency radio-emission of Seyferts along the division line between Seyferts and LINERs in the [NII]-based diagram is dominated by small radio core components, whereas the emission of LINERs is dominated by larger source components. Those could be due to milliarcsecond sized components associated with nuclear jets, compact working points in radio lobes, or very compact wind/shock regions that are bright in the radio domain. This hypothesis is consistent with LINERs' radio morphology.
 \item The spectral index flattening seems to be significant for sources that are in the middle of a galaxy evolutionary sequence that starts with low-metallicity star-forming galaxies and ends with LINERs and passive galaxies, after morphological and color transformation. In this respect, nuclear activity triggering AGN feedback may have a role in first enhancing (in spectroscopically-classified star-forming galaxies) and then shutting down star formation in the hosts, and building up the galaxy color bimodality.
 \item Our results are consistent with previous findings that star-forming galaxies, composites, Seyferts and LINERs show progressively older stellar populations \citep{Vitale2013}, redder colors \citep [e.g.][]{Schawinski2007,Schawinski2009} and higher black hole masses, and with radio-loud AGN being two earlier and later stages of massive galaxy build-up \citep{Hardcastle2006,Smolcic2009}, corresponding to Seyferts and LINERs. The presence of merging and post-merging features in galaxies belonging to the SFGs branch strengthens the scenario of merging-triggered star formation at intermediate redshifts, and sets it at the beginning of the evolutionary sequence.
\end{itemize}
This trend, however, has to be confirmed by follow-up observations extending to lower radio flux densities ($<100$ mJy) on a larger galaxy sample. Future high-resolution imaging will help further constrain the nature of radio-emitters along the flattening sequence in the [NII]-based diagram, and study AGN-feedback in sources that are just leaving the star-forming galaxy branch.

\begin{acknowledgements} 
M. Vitale is supported by the International Max-Planck Research School (IMPRS) for Astronomy and Astrophysics at the Universities of Bonn and Cologne. 
This work is partly based on observations with the 100-m telescope of the MPIfR (Max-Planck-Institut fuer Radioastronomie) at Effelsberg. The authors would like to thank Alex Kraus for support and help during observations and data reduction. 
The FIRST Survey is supported in part under the auspices of the Department of Energy by Lawrence Livermore National Laboratory under contract W-7405-ENG-48 and the Institute for Geophysics and Planetary Physics. 
The Sloan Digital Sky Survey is a joint project of the University of Chicago, Fermilab, the Institute for Advanced Study, the Japan Participation Group, Johns Hopkins University, the Max Planck Institute for Astronomy, the Max Planck Institute for Astrophysics, New Mexico State University, Princeton University, the United States Naval Observatory, and the University of Washington. Apache Point Observatory, site of the SDSS, is operated by the Astrophysical Research Consortium. Funding for the project has been provided by the Alfred P. Sloan Foundation, the SDSS member institutions, NASA, the NSF, the Department of Energy, the Japanese Monbukagakusho, and the Max Planck Society. The SDSS Web site is http://www.sdss.org.
This work was supported in part by the Deutsche Forschungsgemeinschaft (DFG) via the Cologne Bonn Graduate School (BCGS),
and via grant SFB 956. We had fruitful discussions with members of the European Union funded COST Action MP0905: Black Holes in a violent Universe and the COST Action MP1104: Polarization as a tool to study the Solar System and beyond.
We received funding from the European Union Seventh Framework Program (FP7/2007-2013) under grant agreement No.312789. 
\end{acknowledgements}

\vspace*{0.5cm}
\bibliographystyle{aa} 
\bibliography{bib} 

\begin{thebibliography}{134}
\expandafter\ifx\csname natexlab\endcsname\relax\def\natexlab#1{#1}\fi

\bibitem[{{Abazajian} {et~al.}(2009){Abazajian}, {Adelman-McCarthy},
  {Ag{\"u}eros}, {Allam}, {Allende Prieto}, {An}, {Anderson}, {Anderson},
  {Annis}, {Bahcall}, \& et~al.}]{Abazajian2009}
{Abazajian}, K.~N., {Adelman-McCarthy}, J.~K., {Ag{\"u}eros}, M.~A., {et~al.}
  2009, \apjs, 182, 543

\bibitem[{{Afonso} {et~al.}(2005){Afonso}, {Georgakakis}, {Almeida}, {Hopkins},
  {Cram}, {Mobasher}, \& {Sullivan}}]{Afonso2005}
{Afonso}, J., {Georgakakis}, A., {Almeida}, C., {et~al.} 2005, \apj, 624, 135

\bibitem[{{Aird} {et~al.}(2012){Aird}, {Coil}, {Moustakas}, {Blanton},
  {Burles}, {Cool}, {Eisenstein}, {Smith}, {Wong}, \& {Zhu}}]{Aird2012}
{Aird}, J., {Coil}, A.~L., {Moustakas}, J., {et~al.} 2012, \apj, 746, 90

\bibitem[{{Aller} \& {Richstone}(2007)}]{Aller2007}
{Aller}, M.~C. \& {Richstone}, D.~O. 2007, \apj, 665, 120

\bibitem[{Baars {et~al.}(1977)Baars, Genzel, Pauliny-Toth, \&
  Witzel}]{1977A&A....61...99B}
Baars, J. W.~M., Genzel, R., Pauliny-Toth, I. I.~K., \& Witzel, A. 1977,
  A{\&}A, 61, 99

\bibitem[{{Baldi} \& {Capetti}(2008)}]{Baldi2008}
{Baldi}, R.~D. \& {Capetti}, A. 2008, \aap, 489, 989

\bibitem[{{Baldry} {et~al.}(2004){Baldry}, {Glazebrook}, {Brinkmann},
  {Ivezi{\'c}}, {Lupton}, {Nichol}, \& {Szalay}}]{Baldry2004}
{Baldry}, I.~K., {Glazebrook}, K., {Brinkmann}, J., {et~al.} 2004, \apj, 600,
  681

\bibitem[{{Baldwin} {et~al.}(1981){Baldwin}, {Phillips}, \&
  {Terlevich}}]{Baldwin1981}
{Baldwin}, J.~A., {Phillips}, M.~M., \& {Terlevich}, R. 1981, \pasp, 93, 5

\bibitem[{{Barger} {et~al.}(2000){Barger}, {Cowie}, \& {Richards}}]{Barger2000}
{Barger}, A.~J., {Cowie}, L.~L., \& {Richards}, E.~A. 2000, \aj, 119, 2092

\bibitem[{{Baum} {et~al.}(1995){Baum}, {Zirbel}, \& {O'Dea}}]{Baum1995}
{Baum}, S.~A., {Zirbel}, E.~L., \& {O'Dea}, C.~P. 1995, \apj, 451, 88

\bibitem[{{Becker} {et~al.}(1995){Becker}, {White}, \& {Helfand}}]{Becker1995}
{Becker}, R.~H., {White}, R.~L., \& {Helfand}, D.~J. 1995, \apj, 450, 559

\bibitem[{{Benson} {et~al.}(2003){Benson}, {Bower}, {Frenk}, {Lacey}, {Baugh},
  \& {Cole}}]{Benson2003}
{Benson}, A.~J., {Bower}, R.~G., {Frenk}, C.~S., {et~al.} 2003, \apj, 599, 38

\bibitem[{{Best} {et~al.}(2005){Best}, {Kauffmann}, {Heckman}, \&
  {Ivezi{\'c}}}]{Best2005}
{Best}, P.~N., {Kauffmann}, G., {Heckman}, T.~M., \& {Ivezi{\'c}}, {\v Z}.
  2005, \mnras, 362, 9

\bibitem[{{Bower} {et~al.}(2006){Bower}, {Benson}, {Malbon}, {Helly}, {Frenk},
  {Baugh}, {Cole}, \& {Lacey}}]{Bower2006}
{Bower}, R.~G., {Benson}, A.~J., {Malbon}, R., {et~al.} 2006, \mnras, 370, 645

\bibitem[{{Bruzual} \& {Charlot}(2003)}]{BruzualCharlot2003}
{Bruzual}, G. \& {Charlot}, S. 2003, \mnras, 344, 1000

\bibitem[{{Buttiglione} {et~al.}(2010){Buttiglione}, {Capetti}, {Celotti},
  {Axon}, {Chiaberge}, {Macchetto}, \& {Sparks}}]{Buttiglione2010}
{Buttiglione}, S., {Capetti}, A., {Celotti}, A., {et~al.} 2010, \aap, 509, A6

\bibitem[{{Cardamone} {et~al.}(2009){Cardamone}, {Schawinski}, {Sarzi},
  {Bamford}, {Bennert}, {Urry}, {Lintott}, {Keel}, {Parejko}, {Nichol},
  {Thomas}, {Andreescu}, {Murray}, {Raddick}, {Slosar}, {Szalay}, \&
  {Vandenberg}}]{Cardamone2009}
{Cardamone}, C., {Schawinski}, K., {Sarzi}, M., {et~al.} 2009, \mnras, 399,
  1191

\bibitem[{{Cattaneo} {et~al.}(2005){Cattaneo}, {Blaizot}, {Devriendt}, \&
  {Guiderdoni}}]{Cattaneo2005}
{Cattaneo}, A., {Blaizot}, J., {Devriendt}, J., \& {Guiderdoni}, B. 2005,
  \mnras, 364, 407

\bibitem[{{Cattaneo} {et~al.}(2009){Cattaneo}, {Faber}, {Binney}, {Dekel},
  {Kormendy}, {Mushotzky}, {Babul}, {Best}, {Br{\"u}ggen}, {Fabian}, {Frenk},
  {Khalatyan}, {Netzer}, {Mahdavi}, {Silk}, {Steinmetz}, \&
  {Wisotzki}}]{Cattaneo2009}
{Cattaneo}, A., {Faber}, S.~M., {Binney}, J., {et~al.} 2009, \nat, 460, 213

\bibitem[{{Cid Fernandes} {et~al.}(2011){Cid Fernandes}, {Stasi{\'n}ska},
  {Mateus}, \& {Vale Asari}}]{CidFernandes2011}
{Cid Fernandes}, R., {Stasi{\'n}ska}, G., {Mateus}, A., \& {Vale Asari}, N.
  2011, \mnras, 413, 1687

\bibitem[{{Cid Fernandes} {et~al.}(2010){Cid Fernandes}, {Stasi{\'n}ska},
  {Schlickmann}, {Mateus}, {Vale Asari}, {Schoenell}, \&
  {Sodr{\'e}}}]{CidFernandes2010}
{Cid Fernandes}, R., {Stasi{\'n}ska}, G., {Schlickmann}, M.~S., {et~al.} 2010,
  \mnras, 403, 1036

\bibitem[{{Colina} \& {de Juan}(1995)}]{Colina1995}
{Colina}, L. \& {de Juan}, L. 1995, \apj, 448, 548

\bibitem[{{Condon}(1998)}]{Condon1998b}
{Condon}, J.~J. 1998, in IAU Symposium, Vol. 179, New Horizons from
  Multi-Wavelength Sky Surveys, ed. B.~J. {McLean}, D.~A. {Golombek}, J.~J.~E.
  {Hayes}, \& H.~E. {Payne}, 19

\bibitem[{{Condon} {et~al.}(1998){Condon}, {Cotton}, {Greisen}, {Yin},
  {Perley}, {Taylor}, \& {Broderick}}]{Condon1998}
{Condon}, J.~J., {Cotton}, W.~D., {Greisen}, E.~W., {et~al.} 1998, \aj, 115,
  1693

\bibitem[{{Croton} {et~al.}(2006){Croton}, {Springel}, {White}, {De Lucia},
  {Frenk}, {Gao}, {Jenkins}, {Kauffmann}, {Navarro}, \& {Yoshida}}]{Croton2006}
{Croton}, D.~J., {Springel}, V., {White}, S.~D.~M., {et~al.} 2006, \mnras, 365,
  11

\bibitem[{{Dekel} \& {Silk}(1986)}]{DekelSilk1986}
{Dekel}, A. \& {Silk}, J. 1986, \apj, 303, 39

\bibitem[{{Di Matteo} {et~al.}(2005){Di Matteo}, {Springel}, \&
  {Hernquist}}]{DiMatteo2005}
{Di Matteo}, T., {Springel}, V., \& {Hernquist}, L. 2005, \nat, 433, 604

\bibitem[{{Dopita} {et~al.}(2002){Dopita}, {Kewley}, \&
  {Sutherland}}]{Dopita2002}
{Dopita}, M.~A., {Kewley}, L.~J., \& {Sutherland}, R.~S. 2002, in Revista
  Mexicana de Astronomia y Astrofisica, vol. 27, Vol.~12, Revista Mexicana de
  Astronomia y Astrofisica Conference Series, ed. W.~J. {Henney}, J.~{Franco},
  \& M.~{Martos}, 225--229

\bibitem[{{Eckart} {et~al.}(1989){Eckart}, {Hummel}, \& {Witzel}}]{Eckart1989}
{Eckart}, A., {Hummel}, C.~A., \& {Witzel}, A. 1989, \mnras, 239, 381

\bibitem[{{Fanaroff} \& {Riley}(1974)}]{Fanaroff1974}
{Fanaroff}, B.~L. \& {Riley}, J.~M. 1974, \mnras, 167, 31P

\bibitem[{{Feoli} \& {Mancini}(2009)}]{FeoliMancini2009}
{Feoli}, A. \& {Mancini}, L. 2009, \apj, 703, 1502

\bibitem[{{Ferrarese} \& {Merritt}(2000)}]{Ferrarese2000}
{Ferrarese}, L. \& {Merritt}, D. 2000, \apjl, 539, L9

\bibitem[{Fuhrmann {et~al.}(2008)Fuhrmann, Krichbaum, Witzel, Kraus, Britzen,
  Ber~nhart, Impellizzeri, Agudo, Klare, Sohn, Angelakis, Bach, Gab'anyi,
  K{\"o}rding, Pagels, Zensus, Wagner, Ostorero, Ungerechts, Grewing,
  Tornikoski, Apponi, Ziurys, \& Strom}]{2008A&A...490.1019F}
Fuhrmann, L., Krichbaum, T.~P., Witzel, A., {et~al.} 2008, A{\&}A, 490, 1019

\bibitem[{{Gabor} {et~al.}(2009){Gabor}, {Impey}, {Jahnke}, {Simmons}, {Trump},
  {Koekemoer}, {Brusa}, {Cappelluti}, {Schinnerer}, {Smol{\v c}i{\'c}},
  {Salvato}, {Rhodes}, {Mobasher}, {Capak}, {Massey}, {Leauthaud}, \&
  {Scoville}}]{Gabor2009}
{Gabor}, J.~M., {Impey}, C.~D., {Jahnke}, K., {et~al.} 2009, \apj, 691, 705

\bibitem[{{Gaibler} {et~al.}(2012){Gaibler}, {Khochfar}, {Krause}, \&
  {Silk}}]{Gaibler2012}
{Gaibler}, V., {Khochfar}, S., {Krause}, M., \& {Silk}, J. 2012, \mnras, 425,
  438

\bibitem[{{Gebhardt} {et~al.}(2000){Gebhardt}, {Bender}, {Bower}, {Dressler},
  {Faber}, {Filippenko}, {Green}, {Grillmair}, {Ho}, {Kormendy}, {Lauer},
  {Magorrian}, {Pinkney}, {Richstone}, \& {Tremaine}}]{Gebhardt2000}
{Gebhardt}, K., {Bender}, R., {Bower}, G., {et~al.} 2000, \apjl, 539, L13

\bibitem[{{Georgakakis} {et~al.}(2008){Georgakakis}, {Nandra}, {Yan},
  {Willner}, {Lotz}, {Pierce}, {Cooper}, {Laird}, {Koo}, {Barmby}, {Newman},
  {Primack}, \& {Coil}}]{Georgakakis2008}
{Georgakakis}, A., {Nandra}, K., {Yan}, R., {et~al.} 2008, \mnras, 385, 2049

\bibitem[{{Gonzalez-Serrano} {et~al.}(1993){Gonzalez-Serrano}, {Carballo}, \&
  {Perez-Fournon}}]{Gonzalez-Serrano1993}
{Gonzalez-Serrano}, J.~I., {Carballo}, R., \& {Perez-Fournon}, I. 1993, \aj,
  105, 1710

\bibitem[{{Govoni} {et~al.}(2000){Govoni}, {Falomo}, {Fasano}, \&
  {Scarpa}}]{Govoni2000}
{Govoni}, F., {Falomo}, R., {Fasano}, G., \& {Scarpa}, R. 2000, \aap, 353, 507

\bibitem[{{Granato} {et~al.}(2004){Granato}, {De Zotti}, {Silva}, {Bressan}, \&
  {Danese}}]{Granato2004}
{Granato}, G.~L., {De Zotti}, G., {Silva}, L., {Bressan}, A., \& {Danese}, L.
  2004, \apj, 600, 580

\bibitem[{{Gregorini} {et~al.}(1984){Gregorini}, {Mantovani}, {Eckart},
  {Biermann}, {Witzel}, \& {Kuhr}}]{Gregorini1984}
{Gregorini}, L., {Mantovani}, F., {Eckart}, A., {et~al.} 1984, \aj, 89, 323

\bibitem[{{Groves} {et~al.}(2004{\natexlab{a}}){Groves}, {Dopita}, \&
  {Sutherland}}]{Groves2004}
{Groves}, B.~A., {Dopita}, M.~A., \& {Sutherland}, R.~S. 2004{\natexlab{a}},
  \apjs, 153, 9

\bibitem[{{Groves} {et~al.}(2004{\natexlab{b}}){Groves}, {Dopita}, \&
  {Sutherland}}]{Groves2004b}
{Groves}, B.~A., {Dopita}, M.~A., \& {Sutherland}, R.~S. 2004{\natexlab{b}},
  \apjs, 153, 75

\bibitem[{{G{\"u}ltekin} {et~al.}(2009){G{\"u}ltekin}, {Richstone}, {Gebhardt},
  {Lauer}, {Tremaine}, {Aller}, {Bender}, {Dressler}, {Faber}, {Filippenko},
  {Green}, {Ho}, {Kormendy}, {Magorrian}, {Pinkney}, \&
  {Siopis}}]{Gueltekin2009}
{G{\"u}ltekin}, K., {Richstone}, D.~O., {Gebhardt}, K., {et~al.} 2009, \apj,
  698, 198

\bibitem[{{Haggard} {et~al.}(2010){Haggard}, {Green}, {Anderson}, {Constantin},
  {Aldcroft}, {Kim}, \& {Barkhouse}}]{Haggard2010}
{Haggard}, D., {Green}, P.~J., {Anderson}, S.~F., {et~al.} 2010, \apj, 723,
  1447

\bibitem[{{Hamann} {et~al.}(2002){Hamann}, {Korista}, {Ferland}, {Warner}, \&
  {Baldwin}}]{Hamann2002}
{Hamann}, F., {Korista}, K.~T., {Ferland}, G.~J., {Warner}, C., \& {Baldwin},
  J. 2002, \apj, 564, 592

\bibitem[{{Hardcastle} {et~al.}(2006){Hardcastle}, {Evans}, \&
  {Croston}}]{Hardcastle2006}
{Hardcastle}, M.~J., {Evans}, D.~A., \& {Croston}, J.~H. 2006, \mnras, 370,
  1893

\bibitem[{{Harrison} {et~al.}(2012){Harrison}, {Alexander}, {Mullaney},
  {Altieri}, {Coia}, {Charmandaris}, {Daddi}, {Dannerbauer}, {Dasyra}, {Del
  Moro}, {Dickinson}, {Hickox}, {Ivison}, {Kartaltepe}, {Le Floc'h}, {Leiton},
  {Magnelli}, {Popesso}, {Rovilos}, {Rosario}, \& {Swinbank}}]{Harrison2012}
{Harrison}, C.~M., {Alexander}, D.~M., {Mullaney}, J.~R., {et~al.} 2012, \apjl,
  760, L15

\bibitem[{{Heckman}(1980)}]{Heckman1980}
{Heckman}, T.~M. 1980, \aap, 87, 152

\bibitem[{{Heckman} \& {Kauffmann}(2006)}]{HeckmanKauffmann2006}
{Heckman}, T.~M. \& {Kauffmann}, G. 2006, \nar, 50, 677

\bibitem[{{Heckman} {et~al.}(2004){Heckman}, {Kauffmann}, {Brinchmann},
  {Charlot}, {Tremonti}, \& {White}}]{Heckman2004}
{Heckman}, T.~M., {Kauffmann}, G., {Brinchmann}, J., {et~al.} 2004, \apj, 613,
  109

\bibitem[{{Heckman} {et~al.}(1986){Heckman}, {Smith}, {Baum}, {van Breugel},
  {Miley}, {Illingworth}, {Bothun}, \& {Balick}}]{Heckman1986}
{Heckman}, T.~M., {Smith}, E.~P., {Baum}, S.~A., {et~al.} 1986, \apj, 311, 526

\bibitem[{{Hes} {et~al.}(1995){Hes}, {Barthel}, \& {Hoekstra}}]{Hes1995}
{Hes}, R., {Barthel}, P.~D., \& {Hoekstra}, H. 1995, \aap, 303, 8

\bibitem[{{Hickox} {et~al.}(2009){Hickox}, {Jones}, {Forman}, {Murray},
  {Kochanek}, {Eisenstein}, {Jannuzi}, {Dey}, {Brown}, {Stern}, {Eisenhardt},
  {Gorjian}, {Brodwin}, {Narayan}, {Cool}, {Kenter}, {Caldwell}, \&
  {Anderson}}]{Hickox2009}
{Hickox}, R.~C., {Jones}, C., {Forman}, W.~R., {et~al.} 2009, \apj, 696, 891

\bibitem[{{Ho} {et~al.}(2001){Ho}, {Feigelson}, {Townsley}, {Sambruna},
  {Garmire}, {Brandt}, {Filippenko}, {Griffiths}, {Ptak}, \&
  {Sargent}}]{Ho2001}
{Ho}, L.~C., {Feigelson}, E.~D., {Townsley}, L.~K., {et~al.} 2001, \apjl, 549,
  L51

\bibitem[{{Ho} {et~al.}(1997){Ho}, {Filippenko}, \& {Sargent}}]{Ho1997}
{Ho}, L.~C., {Filippenko}, A.~V., \& {Sargent}, W.~L.~W. 1997, \apj, 487, 568

\bibitem[{{Ho} \& {Kim}(2014)}]{Ho2014}
{Ho}, L.~C. \& {Kim}, M. 2014, \apj, 789, 17

\bibitem[{{Hopkins} {et~al.}(2008{\natexlab{a}}){Hopkins}, {Cox}, {Kere{\v s}},
  \& {Hernquist}}]{Hopkins2008b}
{Hopkins}, P.~F., {Cox}, T.~J., {Kere{\v s}}, D., \& {Hernquist}, L.
  2008{\natexlab{a}}, \apjs, 175, 390

\bibitem[{{Hopkins} {et~al.}(2006){Hopkins}, {Hernquist}, {Cox}, {Di Matteo},
  {Robertson}, \& {Springel}}]{Hopkins2006a}
{Hopkins}, P.~F., {Hernquist}, L., {Cox}, T.~J., {et~al.} 2006, \apjs, 163, 1

\bibitem[{{Hopkins} {et~al.}(2008{\natexlab{b}}){Hopkins}, {Hernquist}, {Cox},
  \& {Kere{\v s}}}]{Hopkins2008a}
{Hopkins}, P.~F., {Hernquist}, L., {Cox}, T.~J., \& {Kere{\v s}}, D.
  2008{\natexlab{b}}, \apjs, 175, 356

\bibitem[{{Hopkins} {et~al.}(2007){Hopkins}, {Hernquist}, {Cox}, {Robertson},
  \& {Krause}}]{Hopkins2007}
{Hopkins}, P.~F., {Hernquist}, L., {Cox}, T.~J., {Robertson}, B., \& {Krause},
  E. 2007, \apj, 669, 67

\bibitem[{{Ishibashi} \& {Fabian}(2012)}]{IshibashiFabian2012}
{Ishibashi}, W. \& {Fabian}, A.~C. 2012, \mnras, 427, 2998

\bibitem[{{Ivezi{\'c}} {et~al.}(2002){Ivezi{\'c}}, {Menou}, {Knapp}, {Strauss},
  {Lupton}, {Vanden Berk}, {Richards}, {Tremonti}, {Weinstein}, {Anderson},
  {Bahcall}, {Becker}, {Bernardi}, {Blanton}, {Eisenstein}, {Fan},
  {Finkbeiner}, {Finlator}, {Frieman}, {Gunn}, {Hall}, {Kim}, {Kinkhabwala},
  {Narayanan}, {Rockosi}, {Schlegel}, {Schneider}, {Strateva}, {SubbaRao},
  {Thakar}, {Voges}, {White}, {Yanny}, {Brinkmann}, {Doi}, {Fukugita},
  {Hennessy}, {Munn}, {Nichol}, \& {York}}]{Ivezic2002}
{Ivezi{\'c}}, {\v Z}., {Menou}, K., {Knapp}, G.~R., {et~al.} 2002, \aj, 124,
  2364

\bibitem[{{Ivison} {et~al.}(2012){Ivison}, {Smail}, {Amblard}, {Arumugam}, {De
  Breuck}, {Emonts}, {Feain}, {Greve}, {Haas}, {Ibar}, {Jarvis}, {Kov{\'a}cs},
  {Lehnert}, {Nesvadba}, {R{\"o}ttgering}, {Seymour}, \&
  {Wylezalek}}]{Ivison2012}
{Ivison}, R.~J., {Smail}, I., {Amblard}, A., {et~al.} 2012, \mnras, 425, 1320

\bibitem[{{Kauffmann} {et~al.}(2003{\natexlab{a}}){Kauffmann}, {Heckman},
  {Tremonti}, {Brinchmann}, {Charlot}, {White}, {Ridgway}, {Brinkmann},
  {Fukugita}, {Hall}, {Ivezi{\'c}}, {Richards}, \&
  {Schneider}}]{Kauffmann2003b}
{Kauffmann}, G., {Heckman}, T.~M., {Tremonti}, C., {et~al.} 2003{\natexlab{a}},
  \mnras, 346, 1055

\bibitem[{{Kauffmann} {et~al.}(2003{\natexlab{b}}){Kauffmann}, {Heckman},
  {White}, {Charlot}, {Tremonti}, {Brinchmann}, {Bruzual}, {Peng}, {Seibert},
  {Bernardi}, {Blanton}, {Brinkmann}, {Castander}, {Cs{\'a}bai}, {Fukugita},
  {Ivezic}, {Munn}, {Nichol}, {Padmanabhan}, {Thakar}, {Weinberg}, \&
  {York}}]{Kauffmann2003a}
{Kauffmann}, G., {Heckman}, T.~M., {White}, S.~D.~M., {et~al.}
  2003{\natexlab{b}}, \mnras, 341, 33

\bibitem[{{Kaviraj} {et~al.}(2007){Kaviraj}, {Kirkby}, {Silk}, \&
  {Sarzi}}]{Kaviraj2007}
{Kaviraj}, S., {Kirkby}, L.~A., {Silk}, J., \& {Sarzi}, M. 2007, \mnras, 382,
  960

\bibitem[{{Kawata} \& {Gibson}(2005)}]{KawataGibson2005}
{Kawata}, D. \& {Gibson}, B.~K. 2005, \mnras, 358, L16

\bibitem[{{Kellermann} \& {Pauliny-Toth}(1981)}]{KellermannPauliny-Toth1981}
{Kellermann}, K.~I. \& {Pauliny-Toth}, I.~I.~K. 1981, \araa, 19, 373

\bibitem[{{Kewley} {et~al.}(2001){Kewley}, {Dopita}, {Sutherland}, {Heisler},
  \& {Trevena}}]{Kewley2001}
{Kewley}, L.~J., {Dopita}, M.~A., {Sutherland}, R.~S., {Heisler}, C.~A., \&
  {Trevena}, J. 2001, \apj, 556, 121

\bibitem[{{Kewley} {et~al.}(2003){Kewley}, {Geller}, \& {Jansen}}]{Kewley2003}
{Kewley}, L.~J., {Geller}, M.~J., \& {Jansen}, R.~A. 2003, in Bulletin of the
  American Astronomical Society, Vol.~35, American Astronomical Society Meeting
  Abstracts, 119.01

\bibitem[{{Kewley} {et~al.}(2006){Kewley}, {Groves}, {Kauffmann}, \&
  {Heckman}}]{Kewley2006}
{Kewley}, L.~J., {Groves}, B., {Kauffmann}, G., \& {Heckman}, T. 2006, \mnras,
  372, 961

\bibitem[{{Khalatyan} {et~al.}(2008){Khalatyan}, {Cattaneo}, {Schramm},
  {Gottl{\"o}ber}, {Steinmetz}, \& {Wisotzki}}]{Khalatyan2008}
{Khalatyan}, A., {Cattaneo}, A., {Schramm}, M., {et~al.} 2008, \mnras, 387, 13

\bibitem[{{Klamer} {et~al.}(2004){Klamer}, {Ekers}, {Sadler}, \&
  {Hunstead}}]{Klamer2004}
{Klamer}, I.~J., {Ekers}, R.~D., {Sadler}, E.~M., \& {Hunstead}, R.~W. 2004,
  \apjl, 612, L97

\bibitem[{{Kormendy} \& {Ho}(2013)}]{Kormendy2013}
{Kormendy}, J. \& {Ho}, L.~C. 2013, \araa, 51, 511

\bibitem[{{Kormendy} \& {Richstone}(1995)}]{KormendyRichstone1995}
{Kormendy}, J. \& {Richstone}, D. 1995, \araa, 33, 581

\bibitem[{{Kozie{\l}-Wierzbowska} \& {Stasi{\'n}ska}(2011)}]{Wierzbowska2011}
{Kozie{\l}-Wierzbowska}, D. \& {Stasi{\'n}ska}, G. 2011, \mnras, 415, 1013

\bibitem[{{Lamareille}(2010)}]{Lamareille2010}
{Lamareille}, F. 2010, \aap, 509, A53

\bibitem[{{Lamareille} {et~al.}(2004){Lamareille}, {Mouhcine}, {Contini},
  {Lewis}, \& {Maddox}}]{Lamareille2004}
{Lamareille}, F., {Mouhcine}, M., {Contini}, T., {Lewis}, I., \& {Maddox}, S.
  2004, \mnras, 350, 396

\bibitem[{{Ledlow} \& {Owen}(1995)}]{LedlowOwen1995}
{Ledlow}, M.~J. \& {Owen}, F.~N. 1995, \aj, 110, 1959

\bibitem[{{Ledlow} {et~al.}(1999){Ledlow}, {Owen}, \& {Keel}}]{Ledlow1999}
{Ledlow}, M.~J., {Owen}, F.~N., \& {Keel}, W.~C. 1999, in IAU Symposium, Vol.
  186, Galaxy Interactions at Low and High Redshift, ed. J.~E. {Barnes} \&
  D.~B. {Sanders}, 359

\bibitem[{{Li} {et~al.}(2007){Li}, {Hernquist}, {Robertson}, {Cox}, {Hopkins},
  {Springel}, {Gao}, {Di Matteo}, {Zentner}, {Jenkins}, \& {Yoshida}}]{Li2007}
{Li}, Y., {Hernquist}, L., {Robertson}, B., {et~al.} 2007, \apj, 665, 187

\bibitem[{{Lotz} {et~al.}(2008){Lotz}, {Davis}, {Faber}, {Guhathakurta},
  {Gwyn}, {Huang}, {Koo}, {Le Floc'h}, {Lin}, {Newman}, {Noeske}, {Papovich},
  {Willmer}, {Coil}, {Conselice}, {Cooper}, {Hopkins}, {Metevier}, {Primack},
  {Rieke}, \& {Weiner}}]{Lotz2008}
{Lotz}, J.~M., {Davis}, M., {Faber}, S.~M., {et~al.} 2008, \apj, 672, 177

\bibitem[{{Machalski} \& {Godlowski}(2000)}]{MachalskiGodlowski2000}
{Machalski}, J. \& {Godlowski}, W. 2000, \aap, 360, 463

\bibitem[{{Magorrian} {et~al.}(1998){Magorrian}, {Tremaine}, {Richstone},
  {Bender}, {Bower}, {Dressler}, {Faber}, {Gebhardt}, {Green}, {Grillmair},
  {Kormendy}, \& {Lauer}}]{Magorrian1998}
{Magorrian}, J., {Tremaine}, S., {Richstone}, D., {et~al.} 1998, \aj, 115, 2285

\bibitem[{{Maiolino} {et~al.}(2012){Maiolino}, {Gallerani}, {Neri}, {Cicone},
  {Ferrara}, {Genzel}, {Lutz}, {Sturm}, {Tacconi}, {Walter}, {Feruglio},
  {Fiore}, \& {Piconcelli}}]{Maiolino2012}
{Maiolino}, R., {Gallerani}, S., {Neri}, R., {et~al.} 2012, \mnras, 425, L66

\bibitem[{{Marconi} \& {Hunt}(2003)}]{Marconi2003}
{Marconi}, A. \& {Hunt}, L.~K. 2003, \apjl, 589, L21

\bibitem[{{Martel} {et~al.}(1999){Martel}, {Baum}, {Sparks}, {Wyckoff},
  {Biretta}, {Golombek}, {Macchetto}, {de Koff}, {McCarthy}, \&
  {Miley}}]{Martel1999}
{Martel}, A.~R., {Baum}, S.~A., {Sparks}, W.~B., {et~al.} 1999, \apjs, 122, 81

\bibitem[{{Moles} {et~al.}(1995){Moles}, {Marquez}, \& {Perez}}]{Moles1995}
{Moles}, M., {Marquez}, I., \& {Perez}, E. 1995, \apj, 438, 604

\bibitem[{{Nandra} {et~al.}(2007){Nandra}, {Georgakakis}, {Willmer}, {Cooper},
  {Croton}, {Davis}, {Faber}, {Koo}, {Laird}, \& {Newman}}]{Nandra2007}
{Nandra}, K., {Georgakakis}, A., {Willmer}, C.~N.~A., {et~al.} 2007, \apjl,
  660, L11

\bibitem[{{Narayanan} {et~al.}(2008){Narayanan}, {Cox}, {Kelly}, {Dav{\'e}},
  {Hernquist}, {Di Matteo}, {Hopkins}, {Kulesa}, {Robertson}, \&
  {Walker}}]{Narayanan2008}
{Narayanan}, D., {Cox}, T.~J., {Kelly}, B., {et~al.} 2008, \apjs, 176, 331

\bibitem[{{Narayanan} {et~al.}(2006){Narayanan}, {Cox}, {Robertson},
  {Dav{\'e}}, {Di Matteo}, {Hernquist}, {Hopkins}, {Kulesa}, \&
  {Walker}}]{Narayanan2006}
{Narayanan}, D., {Cox}, T.~J., {Robertson}, B., {et~al.} 2006, \apjl, 642, L107

\bibitem[{{Norris} {et~al.}(2012){Norris}, {Lenc}, {Roy}, \&
  {Spoon}}]{Norris2012}
{Norris}, R.~P., {Lenc}, E., {Roy}, A.~L., \& {Spoon}, H. 2012, \mnras, 422,
  1453

\bibitem[{{Obri{\'c}} {et~al.}(2006){Obri{\'c}}, {Ivezi{\'c}}, {Best},
  {Lupton}, {Tremonti}, {Brinchmann}, {Ag{\"u}eros}, {Knapp}, {Gunn},
  {Rockosi}, {Schlegel}, {Finkbeiner}, {Ga{\'c}e{\v s}a}, {Smol{\v c}i{\'c}},
  {Anderson}, {Voges}, {Juri{\'c}}, {Siverd}, {Steinhardt}, {Jagoda},
  {Blanton}, \& {Schneider}}]{Obric2006}
{Obri{\'c}}, M., {Ivezi{\'c}}, {\v Z}., {Best}, P.~N., {et~al.} 2006, \mnras,
  370, 1677

\bibitem[{{O'Dea}(1998)}]{O'Dea1998}
{O'Dea}, C.~P. 1998, \pasp, 110, 493

\bibitem[{{OMullane} {et~al.}(2005){OMullane}, {Li}, {Nieto-Santisteban},
  {Szalay}, {Thakar}, \& {Gray}}]{OMullane2005}
{OMullane}, W., {Li}, N., {Nieto-Santisteban}, M., {et~al.} 2005, eprint
  arXiv:cs/0502072

\bibitem[{Ott {et~al.}(1994)Ott, Witzel, Quirrenbach, Krichbaum, Standke,
  Schalinski, \& Hummel}]{1994A&A...284..331O}
Ott, M., Witzel, A., Quirrenbach, A., {et~al.} 1994, A{\&}A, 284, 331

\bibitem[{{Owen} \& {Laing}(1989)}]{Owen1989}
{Owen}, F.~N. \& {Laing}, R.~A. 1989, \mnras, 238, 357

\bibitem[{{Panessa} {et~al.}(2007){Panessa}, {Barcons}, {Bassani}, {Cappi},
  {Carrera}, {Ho}, \& {Pellegrini}}]{Panessa2007}
{Panessa}, F., {Barcons}, X., {Bassani}, L., {et~al.} 2007, \aap, 467, 519

\bibitem[{{Ramos Almeida} {et~al.}(2012){Ramos Almeida}, {Bessiere},
  {Tadhunter}, {P{\'e}rez-Gonz{\'a}lez}, {Barro}, {Inskip}, {Morganti}, {Holt},
  \& {Dicken}}]{Ramos-Almeida2012}
{Ramos Almeida}, C., {Bessiere}, P.~S., {Tadhunter}, C.~N., {et~al.} 2012,
  \mnras, 419, 687

\bibitem[{{Rola} {et~al.}(1997){Rola}, {Terlevich}, \& {Terlevich}}]{Rola1997}
{Rola}, C.~S., {Terlevich}, E., \& {Terlevich}, R.~J. 1997, \mnras, 289, 419

\bibitem[{{Sadler} {et~al.}(2002){Sadler}, {Jackson}, {Cannon}, {McIntyre},
  {Murphy}, {Bland-Hawthorn}, {Bridges}, {Cole}, {Colless}, {Collins}, {Couch},
  {Dalton}, {De Propris}, {Driver}, {Efstathiou}, {Ellis}, {Frenk},
  {Glazebrook}, {Lahav}, {Lewis}, {Lumsden}, {Maddox}, {Madgwick}, {Norberg},
  {Peacock}, {Peterson}, {Sutherland}, \& {Taylor}}]{Sadler2002}
{Sadler}, E.~M., {Jackson}, C.~A., {Cannon}, R.~D., {et~al.} 2002, \mnras, 329,
  227

\bibitem[{{Salim} {et~al.}(2007){Salim}, {Rich}, {Charlot}, {Brinchmann},
  {Johnson}, {Schiminovich}, {Seibert}, {Mallery}, {Heckman}, {Forster},
  {Friedman}, {Martin}, {Morrissey}, {Neff}, {Small}, {Wyder}, {Bianchi},
  {Donas}, {Lee}, {Madore}, {Milliard}, {Szalay}, {Welsh}, \& {Yi}}]{Salim2007}
{Salim}, S., {Rich}, R.~M., {Charlot}, S., {et~al.} 2007, \apjs, 173, 267

\bibitem[{{Schawinski}(2009)}]{Schawinski2009}
{Schawinski}, K. 2009, in American Institute of Physics Conference Series, Vol.
  1201, American Institute of Physics Conference Series, ed. S.~{Heinz} \&
  E.~{Wilcots}, 17--20

\bibitem[{{Schawinski} {et~al.}(2007){Schawinski}, {Thomas}, {Sarzi},
  {Maraston}, {Kaviraj}, {Joo}, {Yi}, \& {Silk}}]{Schawinski2007}
{Schawinski}, K., {Thomas}, D., {Sarzi}, M., {et~al.} 2007, \mnras, 382, 1415

\bibitem[{{Schirmer} {et~al.}(2013){Schirmer}, {Diaz}, {Holhjem}, {Levenson},
  \& {Winge}}]{Schirmer2013}
{Schirmer}, M., {Diaz}, R., {Holhjem}, K., {Levenson}, N.~A., \& {Winge}, C.
  2013, \apj, 763, 60

\bibitem[{{Shulevski} {et~al.}(2012){Shulevski}, {Morganti}, {Oosterloo}, \&
  {Struve}}]{Shulevski2012}
{Shulevski}, A., {Morganti}, R., {Oosterloo}, T., \& {Struve}, C. 2012, \aap,
  545, A91

\bibitem[{{Sikora} {et~al.}(2007){Sikora}, {Stawarz}, \& {Lasota}}]{Sikora2007}
{Sikora}, M., {Stawarz}, {\L}., \& {Lasota}, J.-P. 2007, \apj, 658, 815

\bibitem[{{Silk} \& {Rees}(1998)}]{SilkRees1998}
{Silk}, J. \& {Rees}, M.~J. 1998, \aap, 331, L1

\bibitem[{{Silverman} {et~al.}(2008){Silverman}, {Mainieri}, {Lehmer},
  {Alexander}, {Bauer}, {Bergeron}, {Brandt}, {Gilli}, {Hasinger}, {Schneider},
  {Tozzi}, {Vignali}, {Koekemoer}, {Miyaji}, {Popesso}, {Rosati}, \&
  {Szokoly}}]{Silverman2008}
{Silverman}, J.~D., {Mainieri}, V., {Lehmer}, B.~D., {et~al.} 2008, \apj, 675,
  1025

\bibitem[{{Smith} \& {Heckman}(1989)}]{SmithHeckman1989}
{Smith}, E.~P. \& {Heckman}, T.~M. 1989, \apj, 341, 658

\bibitem[{{Smol{\v c}i{\'c}}(2009)}]{Smolcic2009}
{Smol{\v c}i{\'c}}, V. 2009, \apjl, 699, L43

\bibitem[{{Springel} {et~al.}(2005){Springel}, {Di Matteo}, \&
  {Hernquist}}]{Springel2005}
{Springel}, V., {Di Matteo}, T., \& {Hernquist}, L. 2005, \mnras, 361, 776

\bibitem[{{Stasi{\'n}ska} {et~al.}(2006){Stasi{\'n}ska}, {Cid Fernandes},
  {Mateus}, {Sodr{\'e}}, \& {Asari}}]{Stasinska2006}
{Stasi{\'n}ska}, G., {Cid Fernandes}, R., {Mateus}, A., {Sodr{\'e}}, L., \&
  {Asari}, N.~V. 2006, \mnras, 371, 972

\bibitem[{{Stasi{\'n}ska} {et~al.}(2008){Stasi{\'n}ska}, {Vale Asari}, {Cid
  Fernandes}, {Gomes}, {Schlickmann}, {Mateus}, {Schoenell}, {Sodr{\'e}}, \&
  {Seagal Collaboration}}]{Stasinska2008}
{Stasi{\'n}ska}, G., {Vale Asari}, N., {Cid Fernandes}, R., {et~al.} 2008,
  \mnras, 391, L29

\bibitem[{{Stoughton} {et~al.}(2002){Stoughton}, {Lupton}, {Bernardi},
  {Blanton}, {Burles}, {Castander}, {Connolly}, {Eisenstein}, {Frieman},
  {Hennessy}, {Hindsley}, {Ivezi{\'c}}, {Kent}, {Kunszt}, {Lee}, {Meiksin},
  {Munn}, {Newberg}, {Nichol}, {Nicinski}, {Pier}, {Richards}, {Richmond},
  {Schlegel}, {Smith}, {Strauss}, {SubbaRao}, {Szalay}, {Thakar}, {Tucker},
  {Vanden Berk}, {Yanny}, {Adelman}, {Anderson}, {Anderson}, {Annis},
  {Bahcall}, {Bakken}, {Bartelmann}, {Bastian}, {Bauer}, {Berman},
  {B{\"o}hringer}, {Boroski}, {Bracker}, {Briegel}, {Briggs}, {Brinkmann},
  {Brunner}, {Carey}, {Carr}, {Chen}, {Christian}, {Colestock}, {Crocker},
  {Csabai}, {Czarapata}, {Dalcanton}, {Davidsen}, {Davis}, {Dehnen},
  {Dodelson}, {Doi}, {Dombeck}, {Donahue}, {Ellman}, {Elms}, {Evans}, {Eyer},
  {Fan}, {Federwitz}, {Friedman}, {Fukugita}, {Gal}, {Gillespie}, {Glazebrook},
  {Gray}, {Grebel}, {Greenawalt}, {Greene}, {Gunn}, {de Haas}, {Haiman},
  {Haldeman}, {Hall}, {Hamabe}, {Hansen}, {Harris}, {Harris}, {Harvanek},
  {Hawley}, {Hayes}, {Heckman}, {Helmi}, {Henden}, {Hogan}, {Hogg}, {Holmgren},
  {Holtzman}, {Huang}, {Hull}, {Ichikawa}, {Ichikawa}, {Johnston}, {Kauffmann},
  {Kim}, {Kimball}, {Kinney}, {Klaene}, {Kleinman}, {Klypin}, {Knapp},
  {Korienek}, {Krolik}, {Kron}, {Krzesi{\'n}ski}, {Lamb}, {Leger},
  {Limmongkol}, {Lindenmeyer}, {Long}, {Loomis}, {Loveday}, {MacKinnon},
  {Mannery}, {Mantsch}, {Margon}, {McGehee}, {McKay}, {McLean}, {Menou},
  {Merelli}, {Mo}, {Monet}, {Nakamura}, {Narayanan}, {Nash}, {Neilsen},
  {Newman}, {Nitta}, {Odenkirchen}, {Okada}, {Okamura}, {Ostriker}, {Owen},
  {Pauls}, {Peoples}, {Peterson}, {Petravick}, {Pope}, {Pordes}, {Postman},
  {Prosapio}, {Quinn}, {Rechenmacher}, {Rivetta}, {Rix}, {Rockosi}, {Rosner},
  {Ruthmansdorfer}, {Sandford}, {Schneider}, {Scranton}, {Sekiguchi}, {Sergey},
  {Sheth}, {Shimasaku}, {Smee}, {Snedden}, {Stebbins}, {Stubbs}, {Szapudi},
  {Szkody}, {Szokoly}, {Tabachnik}, {Tsvetanov}, {Uomoto}, {Vogeley}, {Voges},
  {Waddell}, {Walterbos}, {Wang}, {Watanabe}, {Weinberg}, {White}, {White},
  {Wilhite}, {Wolfe}, {Yasuda}, {York}, {Zehavi}, \& {Zheng}}]{Stoughton2002}
{Stoughton}, C., {Lupton}, R.~H., {Bernardi}, M., {et~al.} 2002, \aj, 123, 485

\bibitem[{{Tadhunter} {et~al.}(2007){Tadhunter}, {Dicken}, {Holt}, {Inskip},
  {Morganti}, {Axon}, {Buchanan}, {Gonz{\'a}lez Delgado}, {Barthel}, \& {van
  Bemmel}}]{Tadhunter2007}
{Tadhunter}, C., {Dicken}, D., {Holt}, J., {et~al.} 2007, \apjl, 661, L13

\bibitem[{{Tadhunter} {et~al.}(2002){Tadhunter}, {Dickson}, {Morganti},
  {Robinson}, {Wills}, {Villar-Martin}, \& {Hughes}}]{Tadhunter2002}
{Tadhunter}, C., {Dickson}, R., {Morganti}, R., {et~al.} 2002, \mnras, 330, 977

\bibitem[{{Toomre} \& {Toomre}(1972)}]{Toomre1972}
{Toomre}, A. \& {Toomre}, J. 1972, \apj, 178, 623

\bibitem[{{Tremaine} {et~al.}(2002){Tremaine}, {Gebhardt}, {Bender}, {Bower},
  {Dressler}, {Faber}, {Filippenko}, {Green}, {Grillmair}, {Ho}, {Kormendy},
  {Lauer}, {Magorrian}, {Pinkney}, \& {Richstone}}]{Tremaine2002}
{Tremaine}, S., {Gebhardt}, K., {Bender}, R., {et~al.} 2002, \apj, 574, 740

\bibitem[{{Tresse} {et~al.}(1996){Tresse}, {Rola}, {Hammer}, {Stasi{\'n}ska},
  {Le Fevre}, {Lilly}, \& {Crampton}}]{Tresse1996}
{Tresse}, L., {Rola}, C., {Hammer}, F., {et~al.} 1996, \mnras, 281, 847

\bibitem[{{Trump} {et~al.}(2012){Trump}, {Weiner}, {Koo}, {Faber}, \&
  {Kocevski}}]{Trump2012}
{Trump}, J.~R., {Weiner}, B.~J., {Koo}, D.~C., {Faber}, S.~M., \& {Kocevski},
  D.~D. 2012, in American Astronomical Society Meeting Abstracts, Vol. 219,
  American Astronomical Society Meeting Abstracts 219, 131.01

\bibitem[{{vanBreugel} {et~al.}(2004){vanBreugel}, {Fragile}, {Croft}, {de
  Vries}, {Anninos}, \& {Murray}}]{vanBreugel2004}
{vanBreugel}, W., {Fragile}, C., {Croft}, S., {et~al.} 2004, in IAU Symposium,
  Vol. 222, The Interplay Among Black Holes, Stars and ISM in Galactic Nuclei,
  ed. T.~{Storchi-Bergmann}, L.~C. {Ho}, \& H.~R. {Schmitt}, 485--488

\bibitem[{{Veilleux} \& {Osterbrock}(1987)}]{Veilleux1987}
{Veilleux}, S. \& {Osterbrock}, D.~E. 1987, \apjs, 63, 295

\bibitem[{{Vitale} {et~al.}(2013){Vitale}, {Mignoli}, {Cimatti}, {Lilly},
  {Carollo}, {Contini}, {Kneib}, {Le Fevre}, {Mainieri}, {Renzini},
  {Scodeggio}, {Zamorani}, {Bardelli}, {Barnes}, {Bolzonella}, {Bongiorno},
  {Bordoloi}, {Bschorr}, {Cappi}, {Caputi}, {Coppa}, {Cucciati}, {de la Torre},
  {de Ravel}, {Franzetti}, {Garilli}, {Iovino}, {Kampczyk}, {Knobel},
  {Koekemoer}, {Kova{\v c}}, {Lamareille}, {Le Borgne}, {Le Brun},
  {L{\'o}pez-Sanjuan}, {Maier}, {McCracken}, {Moresco}, {Nair}, {Oesch},
  {Pello}, {Peng}, {P{\'e}rez Montero}, {Pozzetti}, {Presotto}, {Silverman},
  {Tanaka}, {Tasca}, {Tresse}, {Vergani}, {Welikala}, \& {Zucca}}]{Vitale2013}
{Vitale}, M., {Mignoli}, M., {Cimatti}, A., {et~al.} 2013, \aap, 556, A11

\bibitem[{{Vitale} {et~al.}(2012){Vitale}, {Zuther},
  {Garc{\'{\i}}a-Mar{\'{\i}}n}, {Eckart}, {Bremer}, {Valencia-S.}, \&
  {Zensus}}]{Vitale2012}
{Vitale}, M., {Zuther}, J., {Garc{\'{\i}}a-Mar{\'{\i}}n}, M., {et~al.} 2012,
  \aap, 546, A17

\bibitem[{{Wills} {et~al.}(2004){Wills}, {Morganti}, {Tadhunter}, {Robinson},
  \& {Villar-Martin}}]{Wills2004}
{Wills}, K.~A., {Morganti}, R., {Tadhunter}, C.~N., {Robinson}, T.~G., \&
  {Villar-Martin}, M. 2004, \mnras, 347, 771

\bibitem[{{Wills} {et~al.}(2002){Wills}, {Tadhunter}, {Robinson}, \&
  {Morganti}}]{Wills2002}
{Wills}, K.~A., {Tadhunter}, C.~N., {Robinson}, T.~G., \& {Morganti}, R. 2002,
  \mnras, 333, 211

\bibitem[{{Witzel} {et~al.}(1979){Witzel}, {Pauliny-Toth}, {Nauber}, \&
  {Schmidt}}]{Witzel1979}
{Witzel}, A., {Pauliny-Toth}, I.~I.~K., {Nauber}, U., \& {Schmidt}, J. 1979,
  \aj, 84, 942

\bibitem[{{York} {et~al.}(2000){York}, {Adelman}, {Anderson}, {Anderson},
  {Annis}, {Bahcall}, {Bakken}, {Barkhouser}, {Bastian}, {Berman}, {Boroski},
  {Bracker}, {Briegel}, {Briggs}, {Brinkmann}, {Brunner}, {Burles}, {Carey},
  {Carr}, {Castander}, {Chen}, {Colestock}, {Connolly}, {Crocker}, {Csabai},
  {Czarapata}, {Davis}, {Doi}, {Dombeck}, {Eisenstein}, {Ellman}, {Elms},
  {Evans}, {Fan}, {Federwitz}, {Fiscelli}, {Friedman}, {Frieman}, {Fukugita},
  {Gillespie}, {Gunn}, {Gurbani}, {de Haas}, {Haldeman}, {Harris}, {Hayes},
  {Heckman}, {Hennessy}, {Hindsley}, {Holm}, {Holmgren}, {Huang}, {Hull},
  {Husby}, {Ichikawa}, {Ichikawa}, {Ivezi{\'c}}, {Kent}, {Kim}, {Kinney},
  {Klaene}, {Kleinman}, {Kleinman}, {Knapp}, {Korienek}, {Kron}, {Kunszt},
  {Lamb}, {Lee}, {Leger}, {Limmongkol}, {Lindenmeyer}, {Long}, {Loomis},
  {Loveday}, {Lucinio}, {Lupton}, {MacKinnon}, {Mannery}, {Mantsch}, {Margon},
  {McGehee}, {McKay}, {Meiksin}, {Merelli}, {Monet}, {Munn}, {Narayanan},
  {Nash}, {Neilsen}, {Neswold}, {Newberg}, {Nichol}, {Nicinski}, {Nonino},
  {Okada}, {Okamura}, {Ostriker}, {Owen}, {Pauls}, {Peoples}, {Peterson},
  {Petravick}, {Pier}, {Pope}, {Pordes}, {Prosapio}, {Rechenmacher}, {Quinn},
  {Richards}, {Richmond}, {Rivetta}, {Rockosi}, {Ruthmansdorfer}, {Sandford},
  {Schlegel}, {Schneider}, {Sekiguchi}, {Sergey}, {Shimasaku}, {Siegmund},
  {Smee}, {Smith}, {Snedden}, {Stone}, {Stoughton}, {Strauss}, {Stubbs},
  {SubbaRao}, {Szalay}, {Szapudi}, {Szokoly}, {Thakar}, {Tremonti}, {Tucker},
  {Uomoto}, {Vanden Berk}, {Vogeley}, {Waddell}, {Wang}, {Watanabe},
  {Weinberg}, {Yanny}, {Yasuda}, \& {SDSS Collaboration}}]{York2000}
{York}, D.~G., {Adelman}, J., {Anderson}, Jr., J.~E., {et~al.} 2000, \aj, 120,
  1579

\bibitem[{{Yuan} {et~al.}(2008){Yuan}, {Zhou}, {Komossa}, {Dong}, {Wang}, {Lu},
  \& {Bai}}]{Yuan2008}
{Yuan}, W., {Zhou}, H.~Y., {Komossa}, S., {et~al.} 2008, \apj, 685, 801

\bibitem[{Zijlstra {et~al.}(2008)Zijlstra, van Hoof, \&
  Perley}]{2008ApJ...681.1296Z}
Zijlstra, A.~A., van Hoof, P. A.~M., \& Perley, R.~A. 2008, ApJ, 681, 1296

\bibitem[{{Zinn} {et~al.}(2013){Zinn}, {Middelberg}, {Norris}, \&
  {Dettmar}}]{Zinn2013}
{Zinn}, P.-C., {Middelberg}, E., {Norris}, R.~P., \& {Dettmar}, R.-J. 2013,
  \apj, 774, 66

\bibitem[{{Zirbel}(1996)}]{Zirbel1996}
{Zirbel}, E.~L. 1996, \apj, 473, 713

\end{thebibliography}

\begin{tiny}
\begin{table*} 
\centering
\caption{\label{all_info}Sources observed with the Effelsberg telescope at $10.45$ and $4.85$ GHz. From left to right: redshift (z), coordinates (RA, DEC), integrated flux density at $1.4$ GHz (F$_{[1.4]}$) in mJy, flux density at $10.45$ GHz (F$_{[10.45]}$) in Jys, its error (Err), flux density at $4.85$ GHz (F$_{[4.85]}$) in Jys, its error (Err), optical morphology (M$_{o}$) from SDSS images, radio morphology (M$_{r}$) from FIRST images, activity type (Activity) from NED, and spectral classification according to the [NII]-based diagnostic diagram (NII$_{d}$). Optical morphology: ``E'' stands for elliptical, ``Cl'' for cluster, ``S'' for spiral, ``SB'' for barred spiral, ``PM'' for post merger, ``Comp'' for compact and ``int'' for interacting. Radio morphology: ``PS'' stands for point source, ``NAT'' for narrow angle tales, ``Extend'' for extended source, ``Asym dbl'' for asymmetric double. Activity type: ``r-l'' stands for radio-loud, ``Blaz'' for blazars, ``c'' for 
candidate.}
\begin{tabular}{cccccccccccccc}
\hline
z & RA & DEC & F$_{[1.4]}$ & F$_{[10.45]}$& Err& F$_{[4.85]}$& Err & M$_{o}$ &  M$_{r}$ & Activity & NII$_{d}$\\
\hline
0.298 & 204.38 & 0.5913 & 127.25 & 0.1366 & 0.0015 & 0.1552 & 0.0027  & E         & PS        & BLLAC      &    C \\
0.106 & 176.30 & -2.994 & 108.93 & 0.0875 & 0.0374 & 0.1213 & 0.0281  & E         & PS        & AGN        &    C \\
0.112 & 185.30 & -2.816 & 105.55 & 0.1516 & 0.0036 & 0.1628 & 0.0108  & E         & PS        & AGN        &    S \\
0.182 & 184.48 & -3.623 & 208.86 & 0.0366 & 0.0043 & 0.0711 & 0.0029  & E         & PS        & AGN        &    S \\
0.247 & 261.84 & 55.181 & 149.13 & 0.2200 & 0.0034 & 0.1624 & 0.0030  & E/Cl      & PS        &            &    L \\
0.041 & 222.34 & 63.270 & 2922.8 & 0.4338 & 0.0049 & 0.9249 & 0.0047  & SB0       & PS/FRI?   & Sy2        &    S \\
0.105 & 240.69 & 52.732 & 575.70 & 0.0886 & 0.0078 & 0.1876 & 0.0024  & S0/E      & PS        & Sy1        &    L \\
0.179 & 239.86 & 53.515 & 182.35 & 0.0282 & 0.0096 & 0.0555 & 0.0019  & E         & PS        & Sy2        &    S \\
0.224 & 251.08 & 45.779 & 115.05 & 0.0684 & 0.0008 & 0.1020 & 0.0024  & E         & lobes/FRI?& BLLAC c    &    L \\
0.084 & 181.02 & 2.4118 & 145.27 & 0.0602 & 0.0031 & 0.1196 & 0.0031  & E/S0      & lobes     &            &    C \\
0.050 & 159.13 & 2.3626 & 202.42 & 0.0518 & 0.0000 & 0.1033 & 0.0029  & S/PM      & Double PS & HII        &    C \\
0.370 & 223.41 & 3.9926 & 378.48 & 0.0893 & 0.0035 & 0.1575 & 0.0028  & E         & PS        &            &    L \\
0.086 & 323.38 & -7.213 & 193.21 & 0.0376 & 0.0000 & 0.0597 & 0.0018  & S         & PS        & AGN        &    S \\
0.040 & 119.61 & 37.786 & 225.50 & 0.4315 & 0.0012 & 0.9773 & 0.0069  & S0        & Jets/FRII &            &    L \\
0.131 & 127.91 & 46.133 & 130.74 & 0.0732 & 0.0116 & 0.1008 & 0.0034  & E         & PS        &            &    L \\
0.225 & 143.62 & 3.0959 & 292.08 & 0.0521 & 0.0023 & 0.1071 & 0.0013  & E         & PS        &            &    L \\
0.319 & 119.09 & 35.911 & 423.67 & 0.0907 & 0.0000 & 0.1653 & 0.0027  & E         & PS        &            &    SF\\
0.055 & 129.15 & 44.019 & 139.30 & 0.0394 & 0.0046 & 0.0609 & 0.0007  & S0        & PS        & AGN        &    S \\
0.097 & 155.57 & 0.5139 & 167.36 & 0.0482 & 0.0012 & 0.0824 & 0.0030  & S0        & PS        & AGN        &    L \\
0.096 & 159.02 & 0.1018 & 109.91 & 0.0774 & 0.0019 & 0.1953 & 0.0036  & Scd       & Jets/FRII &            &    L \\
0.187 & 235.82 & 2.5976 & 442.5  & 0.0699 & 0.0035 & 0.1481 & 0.0027  & E         & PS        & Sy2        &    C \\
0.263 & 201.08 & 4.3186 & 155.16 & 0.0179 & 0.0015 & 0.0413 & 0.0011  & E         & PS        & AGN        &    C \\
0.133 & 211.78 & 4.8837 & 109.33 & 0.0347 & 0.0378 & 0.0391 & 0.0002  & E         & PS        & AGN        &    C \\
0.095 & 228.85 & 4.3627 & 138.32 & 0.0381 & 0.0039 & 0.0726 & 0.0030  & S         & PS        & HII        &    SF\\
0.052 & 230.34 & 4.3418 & 155.38 & 0.2662 & 0.0008 & 0.3424 & 0.0048  & E/S0      & Jet?      &            &    C \\
0.190 & 234.65 & 55.428 & 209.83 & 0.0505 & 0.0012 & 0.0902 & 0.0025  & E         & PS        & Sy2        &    S \\
0.149 & 206.32 & -1.940 & 374.64 & 0.0805 & 0.0023 & 0.1432 & 0.0037  & E/S0      & PS        & AGN/r-l    &    C \\
0.239 & 207.65 & -3.345 & 104.48 & 0.0271 & 0.0004 & 0.0453 & 0.0004  & E         & PS?       & AGN        &    S \\
0.166 & 208.09 & -1.946 & 552.19 & 0.0778 & 0.0070 & 0.1797 & 0.0036  & E         & PS        &            &    S \\
0.132 & 214.17 & -2.936 & 101.12 & 0.0392 & 0.0000 & 0.0711 & 0.0033  & E/S0      & PS        & AGN        &    C \\
0.137 & 234.47 & -0.955 & 107.55 & 0.0250 & 0.0034 & 0.0431 & 0.0022  & E         & PS        & Sy2/NLAGN  &    S \\
0.266 & 223.16 & 6.4606 & 291.60 & 0.0805 & 0.0011 & 0.1683 & 0.0027  & S         & PS        &            &    S \\
0.102 & 229.69 & 6.2322 & 210.86 & 0.1348 & 0.0095 & 0.2368 & 0.0034  & S0        & NAT/jet?  &            &    C \\
0.195 & 339.40 & 0.3441 & 115.98 & 0.0346 & 0.0062 & 0.0499 & 0.0004  & E         & PS        & AGN        &    S \\
0.276 & 355.27 & 0.3093 & 428.95 & 0.1915 & 0.0108 & 0.2661 & 0.0004  & S/E       & PS        & NLAGN      &    L \\
0.266 & 132.51 & 40.602 & 118.29 & 0.1200 & 0.0019 & 0.1064 & 0.0025  & E         & PS        & AGN        &    L \\
0.082 & 122.41 & 34.926 & 154.86 & 0.1058 & 0.0004 & 0.1476 & 0.0026  & E/S0      & jet?      & BLLAC      &    L \\
0.172 & 124.00 & 38.070 & 157.25 & 0.1089 & 0.0008 & 0.1746 & 0.0011  & E         & Extend    &            &    L \\
0.143 & 164.65 & 56.469 & 221.38 & 0.1503 & 0.0780 & 0.2098 & 0.0023  & S/E       & PS        & BLLAC      &    SF\\
0.084 & 136.56 & 46.605 & 313.57 & 0.1194 & 0.0089 & 0.1635 & 0.0009  & SO        & PS        & LINER      &    C \\
0.151 & 156.57 & 45.708 & 105.18 & 0.0659 & 0.0046 & 0.0909 & 0.0019  & E         & PS        &            &    L \\
0.178 & 222.33 & 42.350 & 165.63 & 0.0563 & 0.0033 & 0.0620 & 0.0006  & PM/QSO    & PS        & Blaz       &    S \\
0.214 & 162.71 & 7.9477 & 138.52 & 0.0540 & 0.0015 & 0.0861 & 0.0040  & E         & PS        &            &    S \\
0.129 & 141.36 & 7.4449 & 103.06 & 0.0420 & 0.0004 & 0.0587 & 0.0008  & SO        & PS        &            &    C \\
0.117 & 210.21 & 52.268 & 174.49 & 0.0391 & 0.0056 & 0.0817 & 0.0020  & E         & PS        & AGN        &    L \\
0.099 & 218.84 & 50.856 & 140.96 & 0.0380 & 0.0012 & 0.0674 & 0.0004  & E         & PS        & AGN        &    L \\
0.076 & 212.95 & 52.816 & 321.58 & 0.2412 & 0.0034 & 0.4185 & 0.0048  & E         & Extend    &            &    L \\
0.151 & 242.95 & 40.672 & 553.08 & 0.0757 & 0.0004 & 0.1730 & 0.0035  & E         & PS        &            &    L \\
0.079 & 129.81 & 28.844 & 124.72 & 0.1760 & 0.0131 & 0.2445 & 0.0043  & E/SO      & jet       &            &    L \\
0.115 & 133.34 & 9.4622 & 123.72 & 0.0650 & 0.0027 & 0.0927 & 0.0013  & E/SO      & PS        &            &    S \\
0.206 & 188.45 & 50.439 & 243.22 & 0.0966 & 0.0027 & 0.1466 & 0.0030  & E/S0      & Extend    & Blaz       &    C \\
0.206 & 191.78 & 49.005 & 1212.6 & 0.3512 & 0.0005 & 0.5757 & 0.0018  & E         & PS        & Sy2        &    S \\
0.083 & 184.36 & 15.903 & 137.5  & 0.0682 & 0.0171 & 0.0378 & 0.0030  & S         & PS        &            &    S \\
0.152 & 206.18 & 55.050 & 456.95 & 0.1436 & 0.0005 & 0.2132 & 0.0013  & E         & PS        & AGN        &    S \\
0.152 & 214.94 & 54.387 & 583.66 & 0.6297 & 0.0092 & 0.4965 & 0.0053  & E/S0/QSO  & Double PS & BLLAC      &    SF\\
0.078 & 226.84 & 10.312 & 407.85 & 0.1281 & 0.0085 & 0.2146 & 0.0029  & S0,Comp   & PS        &            &    L \\
0.082 & 210.80 & 6.1691 & 125.0  & 0.1338 & 0.0091 & 0.1781 & 0.0029  & E/S0      & jets      &            &    L \\
0.132 & 207.59 & 9.6696 & 300.57 & 0.1904 & 0.0081 & 0.3315 & 0.0039  & E/Cl      & PS        & Blaz/Sy1.9 &    L \\
0.061 & 220.07 & 5.9427 & 133.99 & 0.1144 & 0.0083 & 0.1353 & 0.0029  & S0        & Extend?   &            &    L \\
0.109 & 230.95 & 32.230 & 169.28 & 0.0314 & 0.0058 & 0.0677 & 0.0016  & S0        & PS        &            &    S \\
\hline
\end{tabular}
\end{table*}
\end{tiny}

\begin{tiny}
\begin{table*} 
\centering
\caption{\label{all_info2} Continued.}
\begin{tabular}{cccccccccccc}
\hline
z & RA & DEC & F$_{[1.4]}$ & F$_{[10.45]}$& Err& F$_{[4.85]}$& Err & M$_{o}$ &  M$_{r}$ & Activity & NII$_{d}$\\
\hline
0.135 & 229.28 & 33.889 & 120.38 & 0.0346 & 0.0000 & 0.0616 & 0.0014  & E/QSO2    & Double PS & Sy1        &      S \\
0.148 & 212.42 & 36.071 & 143.24 & 0.0229 & 0.0004 & 0.0495 & 0.0018  & E/int     & PS        &            &      S \\
0.241 & 128.72 & 55.572 & 8254.5 & 2.4487 & 0.1362 & 5.4221 & 0.0388  & E/int/QSO?& PS        & LINER      &      L \\
0.095 & 124.72 & 22.796 & 187.0  & 0.0862 & 0.0050 & 0.1463 & 0.0026  & E         & Asym dbl  &            &      L \\
0.081 & 164.16 & 14.324 & 173.41 & 0.0302 & 0.0008 & 0.0568 & 0.0060  & S/PM      & PS        &            &      S \\
0.112 & 123.34 & 7.5682 & 462.92 & 0.0720 & 0.0012 & 0.1488 & 0.0039  & E/S0,bar  & PS        & Sy1        &      L \\
0.099 & 151.50 & 34.902 & 3399.3 & 0.8897 & 0.0098 & 1.4429 & 0.0141  & S0        & PS        & LINER      &      S \\
0.098 & 121.50 & 19.104 & 113.29 & 0.0514 & 0.0031 & 0.0769 & 0.0003  & S0        & FRII?     &            &      S \\
0.158 & 138.08 & 27.929 & 230.35 & 0.0500 & 0.0012 & 0.0984 & 0.0011  & E         & PS        &            &      L \\
0.062 & 176.84 & 35.018 & 615.09 & 0.2251 & 0.0244 & 0.2232 & 0.0035  & S0/Comp   & Jets      & BLAC?/Sy2  &      S \\
0.230 & 224.08 & 27.919 & 108.12 & 0.0305 & 0.0066 & 0.0480 & 0.0025  & E         & PS        &            &      L \\
0.289 & 212.22 & 30.350 & 332.55 & 0.0681 & 0.0006 & 0.1277 & 0.0023  & E         & PS        &            &      S \\
0.078 & 188.00 & 33.296 & 101.47 & 0.0200 & 0.0050 & 0.0339 & 0.0002  & S         & Extend    &            &      SF\\
0.045 & 208.07 & 31.446 & 3709.1 & 1.0669 & 0.0706 & 1.8531 & 0.0135  & S         & FRI       & LINER      &      L \\
0.091 & 160.12 & 29.966 & 388.36 & 0.0751 & 0.0288 & 0.1174 & 0.0024  & S         & PS        &            &      L \\
0.159 & 212.06 & 25.565 & 589.27 & 0.1661 & 0.0098 & 0.3051 & 0.0043  & E         & Extend    &            &      L \\
0.079 & 249.98 & 11.466 & 163.94 & 0.0536 & 0.0073 & 0.0858 & 0.0030  & S/int?    & PS        &            &      SF\\
0.154 & 128.06 & 18.536 & 874.23 & 0.3816 & 0.0008 & 0.6079 & 0.0015  & E,N gal   & PS        &            &      S \\
0.089 & 127.27 & 17.904 & 231.80 & 0.1405 & 0.0023 & 0.2010 & 0.0028  & E/SO      & Jets      & BLLAC c    &      L \\
0.155 & 179.22 & 26.542 & 116.58 & 0.0834 & 0.0016 & 0.1016 & 0.0022  & SO?,g     & PS        &            &      S \\
0.112 & 170.12 & 27.603 & 156.75 & 0.0411 & 0.0004 & 0.0591 & 0.0026  & E         & PS        &            &      C \\
0.063 & 162.38 & 23.456 & 105.55 & 0.0427 & 0.0027 & 0.0582 & 0.0018  & E/S0      & Extend    &            &      L \\
0.134 & 167.58 & 21.529 & 289.01 & 0.0470 & 0.0024 & 0.0889 & 0.0015  & S/int?    & PS        &            &      S \\
0.349 & 185.52 & 23.193 & 374.16 & 0.0627 & 0.0000 & 0.1222 & 0.0021  & E         & Extend    &            &      L \\
0.250 & 214.47 & 20.668 & 110.83 & 0.0878 & 0.0016 & 0.1175 & 0.0028  & E         & jet?      &            &      S \\
0.135 & 141.02 & 14.172 & 108.25 & 0.0180 & 0.0008 & 0.0330 & 0.0009  & E/S0/QSO  & PS        &            &      L \\
0.214 & 230.31 & 15.202 & 356.98 & 0.0628 & 0.0038 & 0.1121 & 0.0025  & E         & PS        &            &      L \\
0.178 & 166.75 & 18.430 & 159.92 & 0.0376 & 0.0004 & 0.0634 & 0.0017  & S0/int?   & PS        &            &      S \\
0.137 & 169.27 & 20.235 & 117.45 & 0.0641 & 0.0004 & 0.0737 & 0.0029  & S0        & PS        & Sy2/BLLAC? &      SF\\
0.144 & 212.61 & 14.644 & 434.41 & 0.1544 & 0.0103 & 0.2537 & 0.0036  & E/S0      & Extend    &            &      L \\
0.187 & 227.46 & 15.957 & 406.70 & 0.1185 & 0.0058 & 0.2276 & 0.0054  & E         & Double PS &            &      L \\
0.229 & 187.09 & 16.437 & 105.58 & 0.0355 & 0.0066 & 0.0539 & 0.0028  & E         & lobes     &            &      S \\
0.180 & 202.24 & 17.645 & 158.77 & 0.0300 & 0.0012 & 0.0508 & 0.0019  & E/SO      & PS        &            &      S \\
0.238 & 190.46 & 16.556 & 359.38 & 0.0545 & 0.0039 & 0.1104 & 0.0031  & E         & PS        &            &      L \\
0.066 & 131.33 & 11.431 & 168.82 & 0.0512 & 0.0019 & 0.0871 & 0.0023  & S0        & Extend    &            &      L \\
0.212 & 152.48 & 14.031 & 1044.7 & 0.2435 & 0.0194 & 0.4066 & 0.0045  & E         & PS        &            &      L \\
0.163 & 245.13 & 17.665 & 111.44 & 0.0839 & 0.0019 & 0.0918 & 0.0034  & SO/int?   & PS        &            &      S \\
0.101 & 138.00 & 53.343 & 135.64 & 0.0639 & 0.0015 & 0.0932 & 0.0042  & S0/E      & Double PS & NLAGN/Sy2  &      S \\
0.255 & 201.34 & 3.9802 & 113.29 & 0.0290 & 0.0004 & 0.0585 & 0.0027  & E         & PS        & AGN        &      C \\
0.104 & 7.1392 & 0.9197 & 237.22 & 0.0425 & 0.0015 & 0.0806 & 0.0064  & S/S0      & PS        & Sy2/r-l    &      S \\
0.247 & 37.788 & 0.9509 & 109.26 & 0.0295 & 0.0023 & 0.0418 & 0.0017  & E         & PS        &            &      S \\
0.195 & 221.80 & 40.796 & 396.94 & 0.0899 & 0.0020 & 0.1659 & 0.0011  & E         & Extend    &            &      S \\
0.125 & 172.92 & 47.002 & 127.34 & 0.0422 & 0.0006 & 0.0687 & 0.0011  & S0        & PS        &            &      C \\
0.204 & 173.82 & 12.886 & 139.19 & 0.0418 & 0.0008 & 0.0568 & 0.0031  & E         & PS        &            &      L \\
0.058 & 180.83 & 13.325 & 110.27 & 0.0341 & 0.0031 & 0.0597 & 0.0024  & S0        & Double PS?& AGN        &      S \\
0.258 & 217.05 & 39.205 & 239.44 & 0.1214 & 0.0031 & 0.1462 & 0.0022  & S0/QSO?   & Extend?   & Blaz       &      S \\
0.299 & 176.41 & 44.339 & 327.30 & 0.1794 & 0.0002 & 0.2368 & 0.0017  & S0/QSO?   & Extend    & AGN        &      L \\
0.085 & 196.58 & 11.227 & 117.94 & 0.1233 & 0.0000 & 0.1998 & 0.0029  & E0        & lobes     &            &      L \\
0.066 & 199.41 & 41.262 & 248.86 & 0.1537 & 0.0074 & 0.2280 & 0.0001  & S0        & PS        &            &      L \\
0.131 & 185.38 & 8.3622 & 150.13 & 0.0733 & 0.0134 & 0.0964 & 0.0030  & S0        & PS        & BLLac?     &      L \\
0.249 & 191.08 & 40.860 & 367.63 & 0.0672 & 0.0023 & 0.1757 & 0.0014  & S0/QSO?   & PS        & Sy?        &      S \\
0.083 & 190.53 & 9.4793 & 101.15 & 0.0304 & 0.0016 & 0.0690 & 0.0055  & E         & jet?      &            &      L \\
0.168 & 224.40 & 43.993 & 152.63 & 0.0435 & 0.0052 & 0.0821 & 0.0027  & S0        & jet?      &            &      S \\
0.232 & 234.74 & 35.952 & 146.91 & 0.0556 & 0.0016 & 0.0863 & 0.0026  & E         & Extend    &            &      L \\
0.237 & 253.23 & 25.952 & 125.12 & 0.0422 & 0.0014 & 0.0628 & 0.0000  & E         & PS        &            &      C \\
0.153 & 137.47 & 17.857 & 257.0  & 0.0681 & 0.0011 & 0.1155 & 0.0019  & S0        & PS        &            &      L \\
0.130 & 170.95 & 20.281 & 102.18 & 0.0306 & 0.0034 & 0.0486 & 0.0026  & S0        & PS        &            &      S \\
0.104 & 137.97 & 37.403 & 226.92 & 0.1536 & 0.0004 & 0.2813 & 0.0015  & S0        & Extend    &            &      L \\
0.052 & 229.16 & 0.2505 & 755.71 & 1.0338 & 0.0568 & 1.1906 & 0.0108  & S0/Comp   & FRII      & LINER      &      L \\
\hline
\end{tabular}
\end{table*}
\end{tiny}

\begin{appendix}
 \section \newline
Tab. \ref{all_info} contains general information on our sample, including redshift, coordinates, fluxes, morphological and spectral classification.

\begin{figure*}
  \centering
  \includegraphics[width=18.5cm]{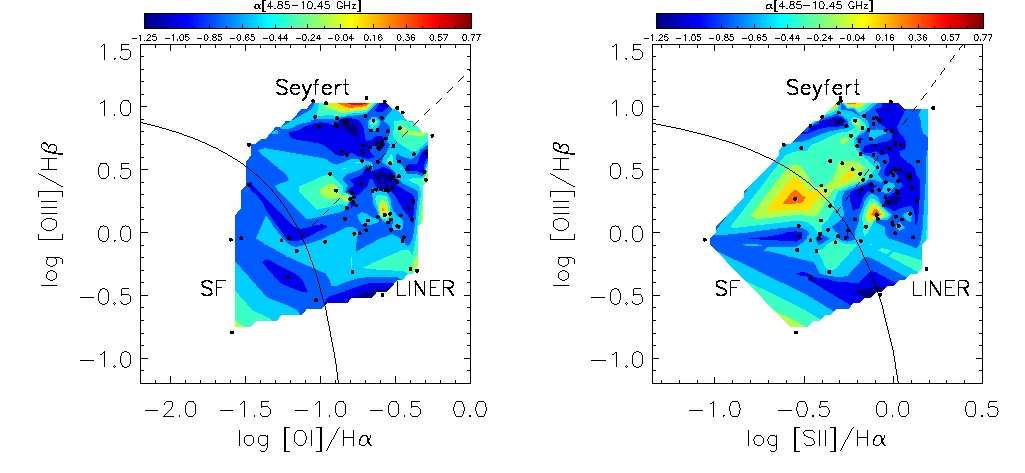}
  \caption{Spectral index distribution in all the low-ionization emission-line diagnostic diagrams. The [O\,{\sc{i}}]-based diagram shows $100$ galaxies, and the [S\,{\sc{ii}}]-based diagram shows $100$.}
 \label{all_dd}
 \end{figure*}
Fig. \ref{all_dd} shows the [O\,{\sc{i}}]- and the [S\,{\sc{ii}}]-based diagnostic diagrams. Here a flattening sequence is less visible, due to the lack of the composite region (Fig. \ref{alpha_BPT}). This is because the \citet{Kewley2006} scheme for the [N\,{\sc{ii}}]-based diagram substantially changes the LINER boundaries, and includes a class for starburst-AGN composites.

\begin{figure*}[!ht]
  \centering
  \includegraphics[width=18.5cm]{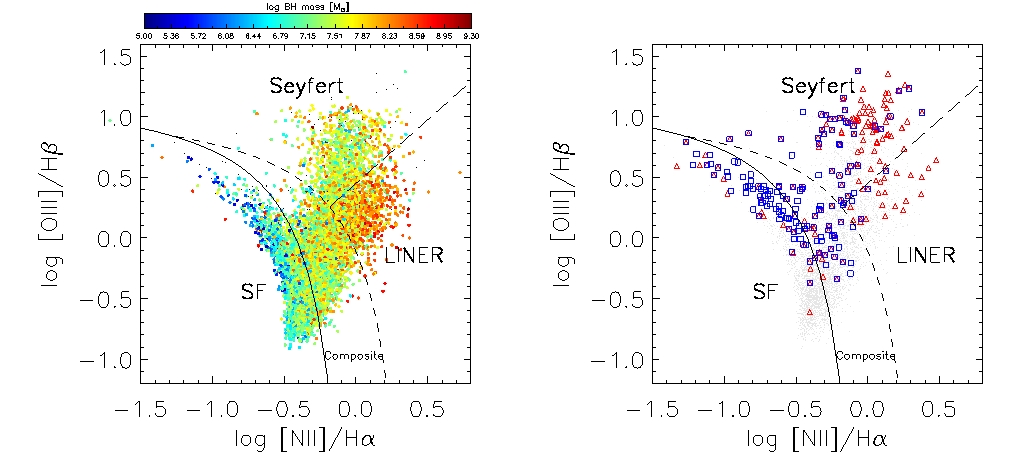}
  \caption{Black hole masses of the objects of the SDSS-FIRST cross-matched sample, represented in the [NII]-based diagram. Left panel: the color bar indicates $M_{\rm BH}$ in solar units. Right panel: grey dots represent the overall black-hole mass distribution as in the left panel; red triangles indicate SDSS sources with velocity dispersion mismeasurements; blue squares indicate sources with $M_{\rm BH}> 10^{9.3}$ $M_{\odot}$ (not flagged $\sigma$ mismeasurements) overestimating black-hole mass measurements}.
 \label{BH_mass_BPT_parent}
 \end{figure*}
The SDSS-FIRST cross-matched sample is composed by $3\,391$ star-forming, $2\,933$ composite, $1\,057$ Seyfert, and $1\,005$ LINER galaxies for which we can trust the black-hole mass estimations. Fig. \ref{BH_mass_BPT_parent} presents the black-hole mass distribution of the objects in this sample. The distribution shows that the black hole mass clearly increases from star-forming galaxies to composites and LINERs. Mean (median) black hole mass values per spectral class are: $7.1$ ($7.2$) for star-forming, $7.5$ ($7.6$) for composite, $7.6$ ($7.6$) for Seyfert, and $8.0$ ($8.1$) for LINER galaxies. The trend shown here is very clear, even though black-hole mass measurements are not direct but derived from the SDSS stellar velocity dispersion. The plot shows a substancial change in the galaxy populations, which is reflected by the Effelsberg sample (Fig. \ref{BH_mass_BPT}).The right panel of Fig. \ref{BH_mass_BPT_parent} presents the overall black-hole mass distribution of the SDSS-FIRST sample 
overimposed with red triangles, which indicate SDSS sources with velocity dispersion mismeasurements ($25$ star-forming, $27$ composite, $102$ Seyfert, and $24$ LINER galaxies), and blue squares, which indicate sources with $M_{\rm BH}> 10^{9.3}$ $M_{\odot}$ ($63$ star-forming, $38$ composite, $43$ Seyfert, and $4$ LINER galaxies). The latter could be again the result of velocity dispersion mismeasurements, not flagged in the SDSS DR7. Some sources are marked as both red triangles and blue squares, indicating that the high black hole masses measured for the objects marked as blue squares could be indeed due to velocity dispersion mismeasurements. It is interesting to note that $57$\% of the red triangles are Seyferts, while $43$\% of the blue squares are star-forming galaxies. Moreover, those Seyferts show especially high  log [O\,{\sc{iii}}]/H$\beta$ ratios ($>1$), and those star-forming galaxies occupy the metal-poor part of the star-forming branch.

\begin{figure*}
  \centering
  \includegraphics[width=9cm]{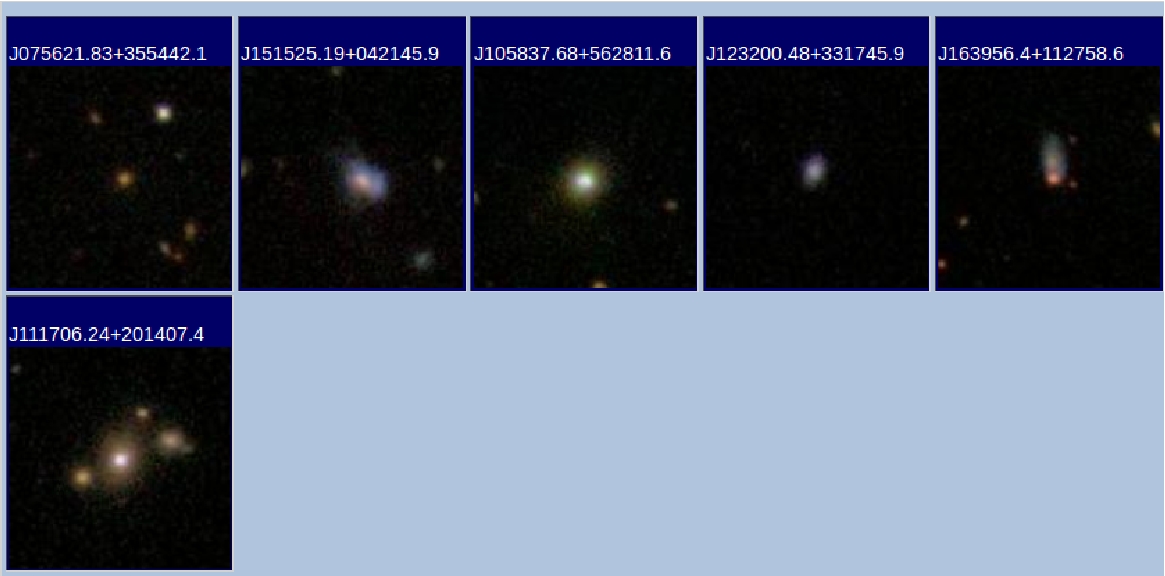}
  \caption{Optical (SDSS) images of star-forming galaxies drawn from our optical-radio sample.}
 \label{sfgs}
 \end{figure*}

\begin{figure*}
  \centering
  \includegraphics[width=9cm]{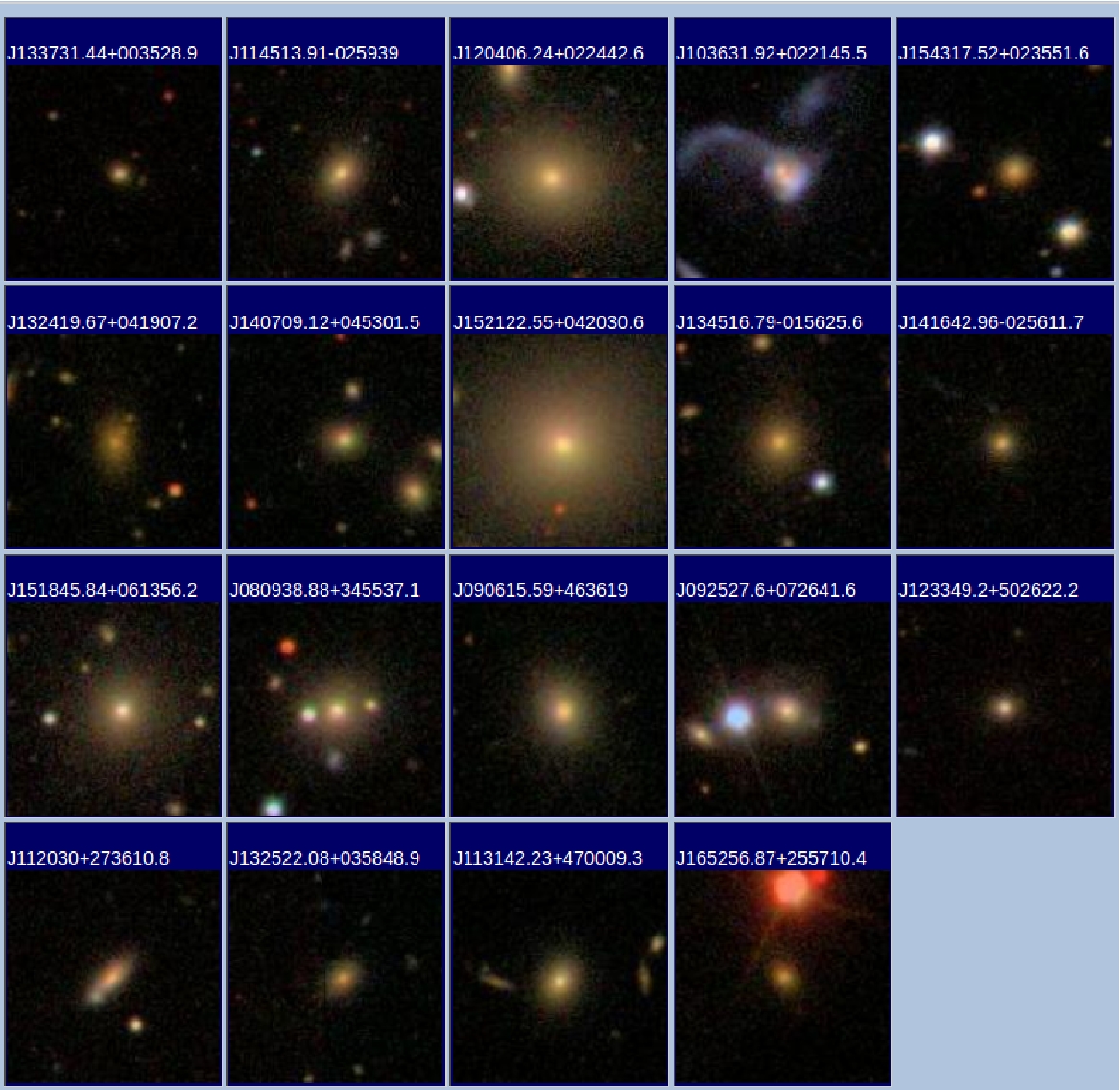}
  \caption{Composite galaxies.}
 \label{composites}
 \end{figure*}

\begin{figure*}
  \centering
  \includegraphics[width=10cm]{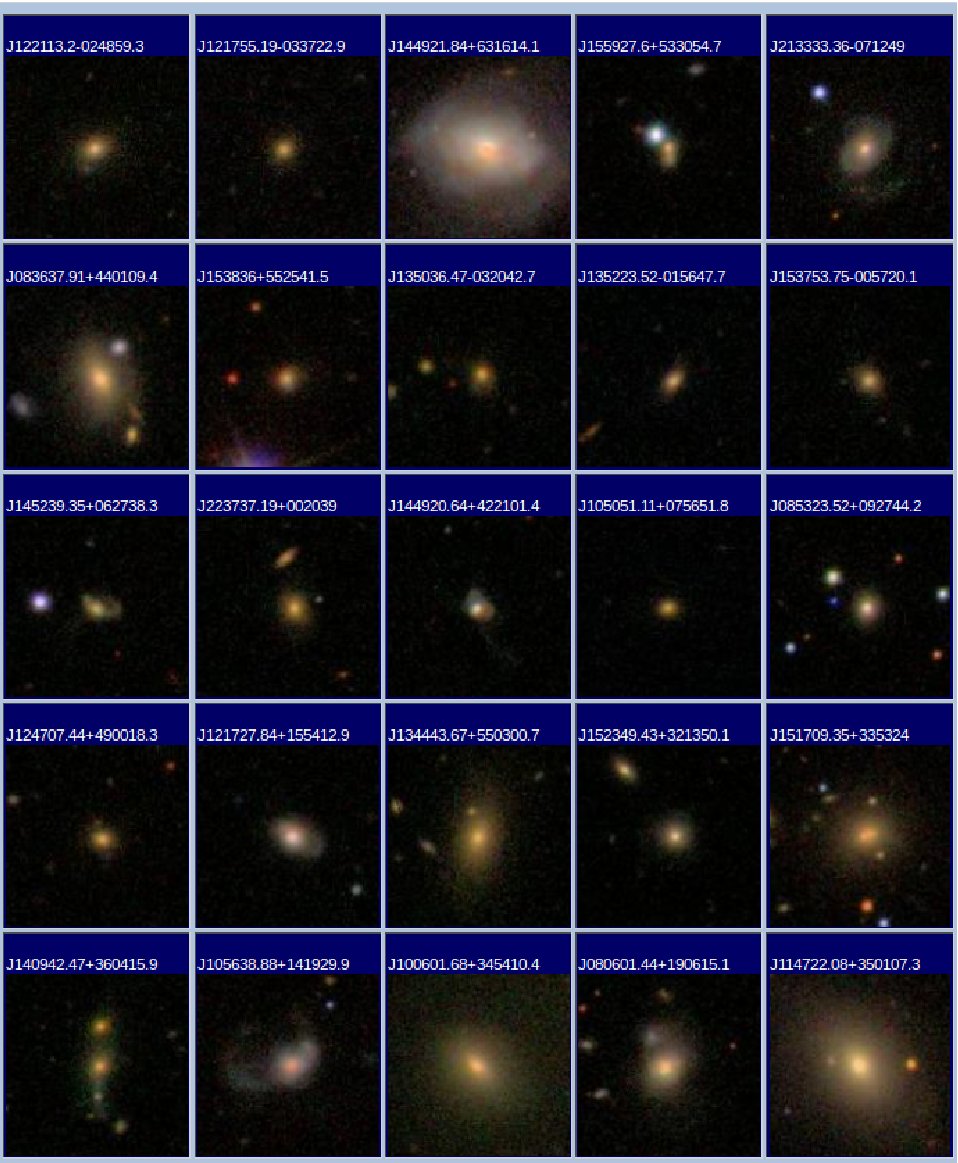}

   \includegraphics[width=10cm]{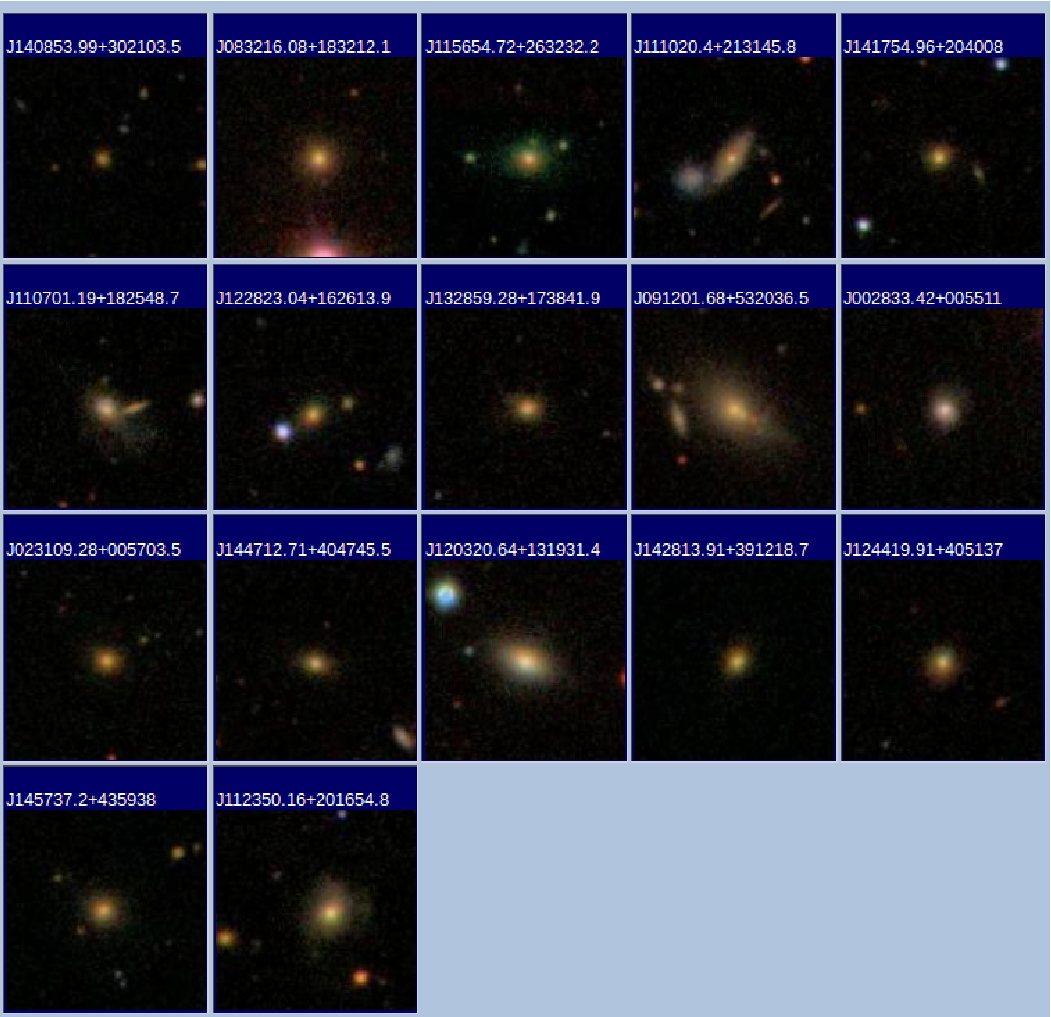}
   \caption{Seyfert galaxies.}

 \label{seyfert}
 \end{figure*}

\begin{figure*}
  \centering
  \includegraphics[width=9cm]{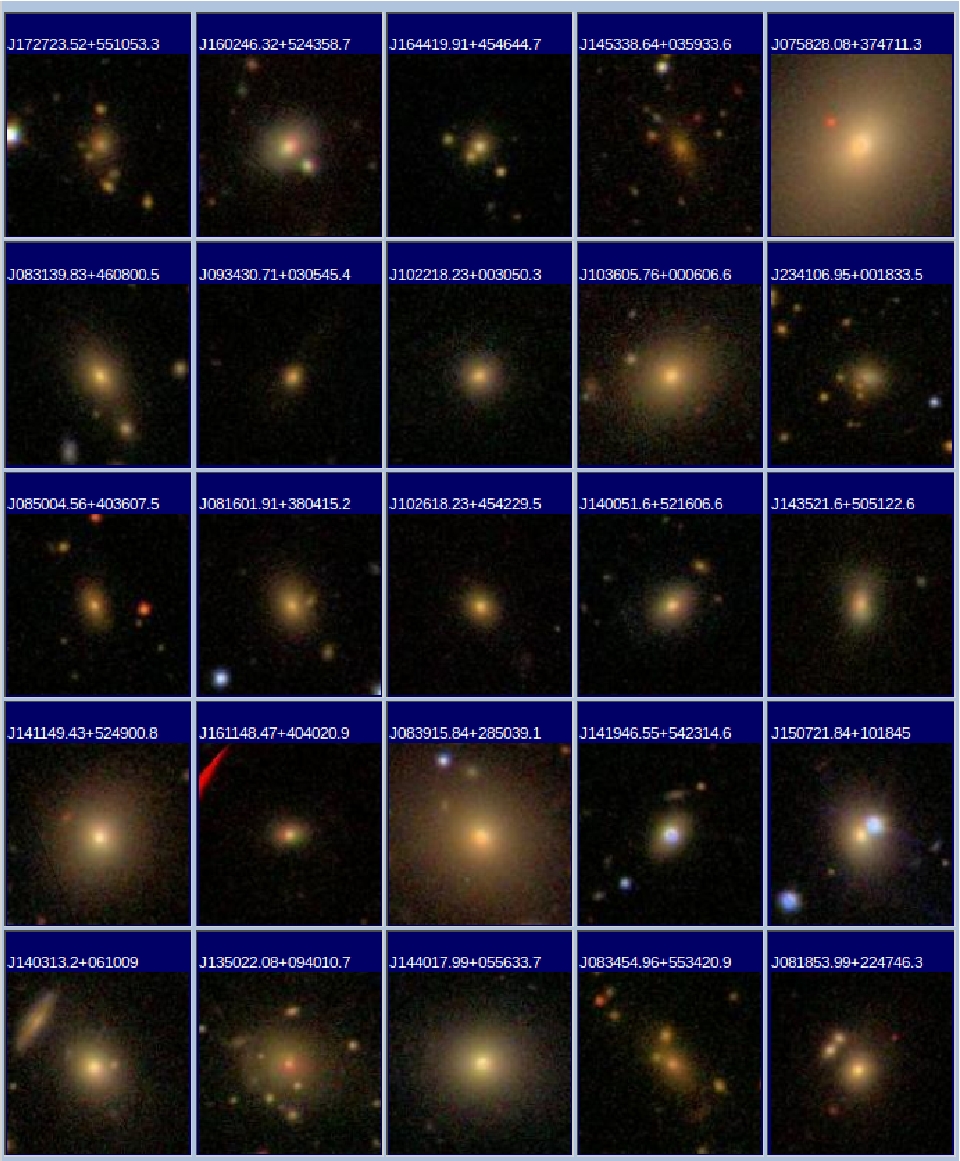}

  \includegraphics[width=9cm]{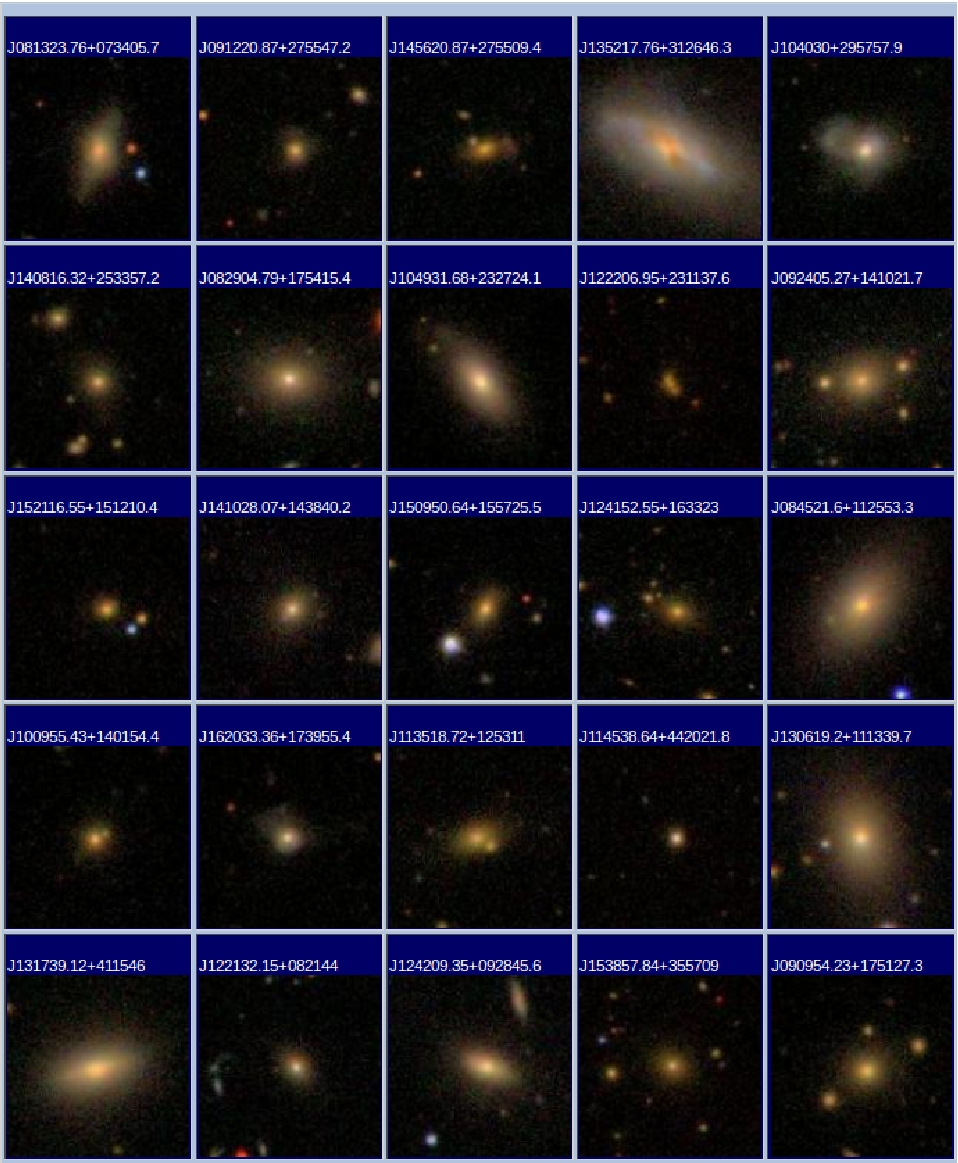}

  \includegraphics[width=4cm]{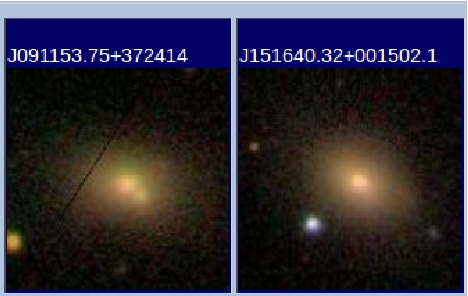}
  \caption{LINER galaxies}
 \label{liners}
 \end{figure*}

The color composite images from SDSS are shown in Fig. \ref{sfgs} to \ref{liners} illustrate the optical morphology of our radio-emitting star-forming, composite, Seyfert and LINER galaxies. Star-forming galaxies are in general blue, and the galaxy color gets progressively redder in the other spectral types. Only a few galaxies are close enough ($z\sim 0.05$) for their morphology to be determined by eye.

\end{appendix}

\end{document}